\documentclass[]{aa}
\usepackage[varg]{txfonts}
\usepackage{lscape}

\begin{document}

   \title{The effects of a background potential in star cluster evolution}

   \subtitle{A delay in the relaxation time-scale and runaway collision processes}

   \author{B. Reinoso
          \inst{1,2}
          \and
          D.R.G. Schleicher\inst{1}
          \and
          M. Fellhauer\inst{1}
          \and
          N.W.C. Leigh\inst{1,3}
          \and
          R.S. Klessen\inst{2,4}
          }

   \institute{Departamento de Astronom\'ia, Facultad Ciencias F\'isicas y Matem\'aticas, Universidad de Concepci\'on, Av. Esteban Iturra s/n Barrio Universitario, Casilla 160-C, Concepci\'on, Chile; \\
              \email{breinoso@udec.cl}
         \and
             Universit\"at Heidelberg, Zentrum f\"ur Astronomie, Institut f\"ur Theoretische Astrophysik, Albert-Ueberle-Str. 2, 69120 Heidelberg, Germany
         \and
          Department of Astrophysics, American Museum of Natural History, New York, NY 10024, USA
         \and
          Universit\"{a}t Heidelberg, Interdisziplin\"{a}res Zentrum f\"{u}r Wissenschaftliches Rechnen, Im Neuenheimer Feld 205, 69120 Heidelberg, Germany
             } 

   \date{Received September 15, 1996; accepted March 16, 1997}

 
  \abstract{Runaway stellar collisions in dense star clusters are invoked to explain the presence of very massive stars or blue stragglers in the center of those systems. This process has also been explored for the first star clusters in the Universe and shown to yield stars that may collapse at some points into an intermediate mass black hole. Although the early evolution of star clusters requires the explicit modeling of the gas out of which the stars form, these calculations would be extremely time-consuming and often the effects of the gas can be accurately treated by including a background potential to account for the extra gravitational force. We apply this approximation to model the early evolution of the first dense star clusters formed in the Universe by performing $N$-body simulations, our goal is to understand how the additional gravitational force affects the growth of a very massive star through stellar mergers in the central parts of the star cluster. Our results show that the background potential increases the velocities of the stars, causing an overall delay in the evolution of the clusters and in the runaway growth of a massive star at the center. The population of binary stars is lower due to the increased kinetic energy of the stars, initially reducing the number of stellar collisions, and we show that relaxation processes are also affected. Despite these effects, the external potential enhances the mass of the merger product by a factor $\sim$2 if the collisions are maintained for long times.}


   \maketitle

\section{Introduction}
The discovery of supermassive black holes (SMBHs) in the first billion years of the Universe \citep{Fan2006,Mortlock2011,Wu2015,Reed2019} has led to the study of formation channels for massive black hole seeds early on ($z\sim 10-20$), following the collapse of molecular-cooling halos ($\sim 10^6$~M$_{\odot}$) or atomic-cooling halos ($\sim 10^8$~M$_\odot$). The direct collapse of protogalactic clouds exposed to a moderate Lyman-Werner flux from neighboring star forming halos produced the most massive seeds ($10^{3-5}$~M$_\odot$) \citep{Wise2019}, and recent numerical simulations suggest that contrary to what was previously thought, the UV flux is not that critical anymore in the case of the efficient merger of fragments \citep{Suazo2019}. Another promising mechanism to form massive seeds is the formation of very massive stars (VMS) in the centers of dense stellar systems through stellar mergers \citep{Fujii2013,Katz2015,Sakurai2017,Reinoso2018}, producing either VMSs with several $10^2$~M$_\odot$ or intermediate mass black holes with $\sim 10^3$~M$_\odot$ and potentially up to $10^4$~M$_\odot$. Such massive seeds could be present in the early Universe if accreting population III~(Pop.~III) protostars can merge before entering the main sequence or if the remaining gas in the cluster can be accreted by the central object \citep{Boekholt2018}. Furthermore, after the formation of a BH in the center of a stellar cluster, additional growth can be expected by tidal disruption events of stars passing close to the black hole \citep{Sakurai2019,Bonetti2020}.

Although most of these previous studies have focused on mergers in second generation star clusters, that is, clusters that formed from molecular clouds that were polluted by stellar winds or supernovae from the first generation of stars \citep{Katz2015,Sakurai2017,Sakurai2019}, we are mostly interested in the very first star clusters of the Universe given the particular properties of these systems and these stars. Fragmentation occurs at higher densities in primordial gas clouds ($10^9$~cm$^{-3}$ or higher) \citep[e.g.,][]{Clark2011b,Greif2011,Greif2012,Smith2011,Smith2012,Latif2013b}, with clusters potentially having half-mass radii of 0.1~pc, especially if dust grains are present in such clouds which trigger fragmentation at high densities \citep[e.g.,][]{Omukai2005,Klessen2012,Bovino2016,Latif2016}.

In those environments, the protostellar radii could also be enhanced due to the rapid accretion expected in primordial or low-metallicity gas, which in turn are a consequence of much higher gas temperatures than at present day star formation, and so the protostellar radii are enhanced up to 300~R$_\odot$ \citep{Stahler1986,Omukai2001,Omukai2003}. Moreover, high accretion rates of $\sim 0.1$~M$_\odot$~yr$^{-1}$ have been reported in several simulations for these protostars \citep{Hosokawa2012,Hosokawa2013,Schleicher2013,Haemmerle2018,Woods2017}, implying $\sim 500$~R$_\odot$ for a 10~M$_\odot$ star and potentially more than 1000~R$_\odot$ for a 100~M$_\odot$ star. While some models already included the effects of an external potential in the evolution of dense star clusters \citep[e.g.,][]{Leigh2013,Leigh2013_2,Leigh2014,Boekholt2018,Sakurai2019} in order to mimic the effects of gas during the embedded phase of the cluster or to account for the dark matter halo, disentangling the effects of the external potential can be tricky because these models also include mass accretion recipes \citep{Boekholt2018}.

In this paper, we present a systematic investigation on how the formation of very massive stars in dense star clusters and the evolution of the clusters themselves are affected when including a background potential in the evolution of the systems. Our work can be used to constraint the parameter space that future, more sophisticated simulations should focus on. Since each physical collision led to a merger in our runs, we use these terms interchangeably. We describe our setup and initial conditions in section~\ref{sec:simulation_setup} and our results are presented in section~\ref{sec:Results}, including the typical evolution of clusters under the influence of a background potential and clusters without a background potential, the number of collisions, the mass of the resulting object, and our model to estimate the number of mergers at different times.
We apply our model to population III (Pop.III) star clusters in Sec.~\ref{sec:implications_for_primordial_clusters}. A discussion about neglected effects and considerations for future research on this subject is given in Sec.~\ref{sec:discussion} and finally the conclusions are presented in Sec.~\ref{sec:conclusions}.

\section{Simulation setup}
\label{sec:simulation_setup}

To understand the impact of the gas potential on the number of mergers that occur in star clusters, we first performed simulations in which we did not include stellar mergers, and we compared the evolution of these systems without an external potential and when placed in the center of a background potential. Once we addressed the effects of the extra force on the evolution of the clusters, we performed simulations that include stellar mergers
in clusters with and without an external potential. To perform the calculations, we used a modified version of NBODY6$^{1}$\citep{Aarseth2000} to treat collisions, where we switched off the stellar evolution package and instead explicitly specify the stellar radii to perform a parameter study. NBODY6 is a fourth order Hermite integrator, and includes a spatial hierarchy to speed up the calculations: This is referred to as the Ahmad-Cohen scheme \citep{AhmadCohen73}. It also includes routines to treat tidal capture and tidal circularization of binary systems \citep{Mardling2001} that we activated in our simulations. Another important routine included is the Kustaanheimo-Stiefel regularization \citep{Kustaanheimo1965}, which is an algorithm that can be used to treat binaries and close two-body encounters more accurately. We performed a total of 344 $N$-body simulations.

\subsection{Simulations without stellar mergers}
\label{sec:sim_no_mergers}
In order to create a controlled experiment to which we can compare our results including stellar mergers, we first investigated the effects produced by the external potential on the evolution of the star clusters in the absence of collisions. We specifically explored how the mass of the external potential affects the core-collapse, the formation of binary systems, and the evaporation of the clusters. We modeled each cluster as a Plummer \citep{Plummer1911} distribution of $N=1\,000$ and $N=10\,000$ equal mass particles with a total cluster mass of $M_{\rm stars}=10^4$~M$_\odot$. Given that we are interested in the overall evolution of the system, and in order to save computational time, we modeled less dense clusters than the ones presented in Sec.~\ref{sec:sim_mergers}, with a virial radius of $R_{\rm v}=1.0$~pc. We then included an analytic background potential that slso follows a Plummer density profile with the same virial radius as the stellar distribution $R_{\rm v,ext}=1.0$~pc and the mass of the potential was varied as $M_{\rm ext}=0.0,0.1,0.5$ and $1.0\ \times$ $M_{\rm stars}$. The clusters start in virial equilibrium. We performed a total of eight simulations that do not include stellar mergers and which are listed in Table~\ref{tab:sim_no_mergers}.

\begin{table*}
\caption{Simulations that do not include stellar mergers} 
\label{tab:sim_no_mergers}
\centering
\begin{tabular}{rrcccr}
\hline \hline
Number & $N$ & $M_{\rm stars}$ [M$_\odot$] & $M_{\rm ext}/M_{\rm stars}$ &$R_{\rm v}$ [pc] & $t_{\rm cc}/t_{\rm cross}$  \\
\hline
1 & $1\,000$ &$10^4$ & 0.0 &1.0 &456\\
2 & $1\,000$ &$10^4$ & 0.1 &1.0 &460\\
3 & $1\,000$ &$10^4$ & 0.5 &1.0 &1\,820\\
4 & $1\,000$ &$10^4$ & 1.0 &1.0 &6\,553\\

5 & $10\,000$ &$10^4$ & 0.0 &1.0 &2\,306\tablefootmark{a}\\
6 & $10\,000$ &$10^4$ & 0.1 &1.0 &3\,377\tablefootmark{a}\\
7 & $10\,000$ &$10^4$ & 0.5 &1.0 &13\,519\tablefootmark{a}\\
8 & $10\,000$ &$10^4$ & 1.0 &1.0 &51\,889\tablefootmark{a}\\
\hline
\end{tabular}
\tablefoot{Summary of simulations that do not include stellar mergers showing the number of particles, the cluster mass, the ratio of external potential to cluster mass, the virial radius and the ratio of core collapse time to crossing time in columns 1,2,3,4,5 and 6 respectively. 
\tablefoottext{a}{Crossing time calculated using Eq.(\ref{eq:tcross}) with $q=M_{\rm ext}/M_{\rm stars}=1.0$.}}

\end{table*}

\subsection{Simulations that include stellar mergers}
\label{sec:sim_mergers}
Our simulations that include stellar mergers aim to model the first stages after the formation of Pop.~III star clusters. We modeled a compact cluster in virial equilibrium by also using a Plummer distribution \citep{Plummer1911} for the stars that are all equal in mass and radius at the beginning of the simulation. We modeled dense clusters with a virial radius of $R_{\rm v}=0.14$ pc and a total mass in stars of $M_{\rm stars}=10^4$~M$_{\odot}$. We then performed the same set of simulations including an external analytic Plummer potential with $M_{\rm ext}=1.0\ \times$~$M_{\rm stars}$ with the same virial radius of $R_{\rm v,ext}=0.14$~pc in order to consider at first order the effects of the gas that remains in the cluster after the formation of the stars. Taking the mass of the external potential into account, the crossing time of the clusters were calculated as:
\begin{equation}
\label{eq:tcross}
t_{\rm cross} = \sqrt{\frac{R_{\rm v}^3}{G M_{\rm stars}}} \frac{1}{1+q},    
\end{equation}
with $q =\frac{M_{\rm ext}}{M_{\rm stars}}$ (see Appendix~\ref{sec:appendix1}); this gives a value of $t_{\rm cross}=0.0078$~Myr and $t_{\rm cross,ext}=0.0039$~Myr for clusters without and with an external potential, respectively.
We investigated how the mass of the final merger product depends on the initial number of stars in the cluster $N$, which we varied as $N=100,500,1\,000$, and $5\,000$ stars. We also varied the initial stellar radius as $R_{\rm star}=20,50,100,200,500,1000$, and $5\,000$~R$_\odot$ using equal radii stars in each run. For each of these configurations, we performed simulations with $M_{\rm ext}=0.0$ and $1.0\times$~$M_{\rm stars}$. Finally, we ran a total of six random simulations per each configuration, which are listed in Table~\ref{tab:sim_mergers}. This gives a total of 336 $N$-body simulations that include stellar mergers.

\begin{table*}
\caption{\label{tab:sim_mergers} Simulations that include stellar mergers}
\centering
\begin{tabular}{rrcccrrr}
\hline \hline
Number & $N$ & $M_{\rm stars}$ [M$_\odot$] & $M_{\rm ext}/M_{\rm stars}$ &$R_{\rm v}$ [pc] & $m_{\rm star}$ [M$_\odot$] & $R_{\rm star}$ [R$_\odot$] & $\overline{\rm M}_{\rm max}$ [M$_\odot$] \tablefootmark{a} \\
\hline
1 & $100$ &$10^4$ & 0.0 &0.14 &100.0&20.0& 350 $\pm$ 164 \\
2 & $100$ &$10^4$ & 0.0 &0.14 &100.0&50.0& 466 $\pm$ 137\\
3 & $100$ &$10^4$ & 0.0 &0.14 &100.0&100.0& 883 $\pm$ 237\\
4 & $100$ &$10^4$ & 0.0 &0.14 &100.0&200.0& 1\,050 $\pm$ 274 \\
5 & $100$ &$10^4$ & 0.0 &0.14 &100.0&500.0& 1\,533 $\pm$ 413\\
6 & $100$ &$10^4$ & 0.0 &0.14 &100.0&1\,000.0& 2\,083 $\pm$ 691\\
7 & $100$ &$10^4$ & 0.0 &0.14 &100.0&5\,000.0& 2\,833 $\pm$ 516\\

8 & $500$ &$10^4$ & 0.0 &0.14 &20.0&20.0& 390 $\pm$ 55 \\
9 & $500$ &$10^4$ & 0.0 &0.14 &20.0&50.0& 633 $\pm$ 120 \\
10 & $500$ &$10^4$ & 0.0 &0.14 &20.0&100.0& 937 $\pm$ 184 \\
11 & $500$ &$10^4$ & 0.0 &0.14 &20.0&200.0& 1\,390 $\pm$ 170 \\
12 & $500$ &$10^4$ & 0.0 &0.14 &20.0&500.0& 2\,056 $\pm$ 204 \\
13 & $500$ &$10^4$ & 0.0 &0.14 &20.0&1\,000.0& 2\,756 $\pm$ 463 \\
14 & $500$ &$10^4$ & 0.0 &0.14 &20.0&5\,000.0& 3\,013 $\pm$ 332 \\

15 & $1\,000$ &$10^4$ & 0.0 &0.14 &10.0&20.0& 430 $\pm$ 143 \\
16 & $1\,000$ &$10^4$ & 0.0 &0.14 &10.0&50.0& 747 $\pm$ 173 \\
17 & $1\,000$ &$10^4$ & 0.0 &0.14 &10.0&100.0& 1\,168 $\pm$ 132 \\
18 & $1\,000$ &$10^4$ & 0.0 &0.14 &10.0&200.0& 1\,748 $\pm$ 188 \\
19 & $1\,000$ &$10^4$ & 0.0 &0.14 &10.0&500.0& 2\,428 $\pm$ 150 \\
20 & $1\,000$ &$10^4$ & 0.0 &0.14 &10.0&1\,000.0& 2\,980 $\pm$ 243 \\
21 & $1\,000$ &$10^4$ & 0.0 &0.14 &10.0&5\,000.0& 3\,217 $\pm$ 162 \\

22 & $5\,000$ &$10^4$ & 0.0 &0.14 &2.0&20.0& 421 $\pm$ 263\\
23 & $5\,000$ &$10^4$ & 0.0 &0.14 &2.0&50.0& 1\,005 $\pm$ 377 \\
24 & $5\,000$ &$10^4$ & 0.0 &0.14 &2.0&100.0& 1\,520 $\pm$ 284 \\
25 & $5\,000$ &$10^4$ & 0.0 &0.14 &2.0&200.0& 1\,912 $\pm$ 750 \\
26 & $5\,000$ &$10^4$ & 0.0 &0.14 &2.0&500.0& 3\,223 $\pm$ 305 \\
27 & $5\,000$ &$10^4$ & 0.0 &0.14 &2.0&1\,000.0& 3\,495 $\pm$ 146 \\
28 & $5\,000$ &$10^4$ & 0.0 &0.14 &2.0&5\,000.0& 4\,256 $\pm$ 366 \\

29 & $100$ &$10^4$ & 1.0 &0.14 &100.0&20.0& 683 $\pm$ 216 \\
30 & $100$ &$10^4$ & 1.0 &0.14 &100.0&50.0& 983 $\pm$ 293 \\
31 & $100$ &$10^4$ & 1.0 &0.14 &100.0&100.0& 1\,583 $\pm$ 581 \\
32 & $100$ &$10^4$ & 1.0 &0.14 &100.0&200.0& 2\,183 $\pm$ 382 \\
33 & $100$ &$10^4$ & 1.0 &0.14 &100.0&500.0& 2\,800 $\pm$ 316 \\
34 & $100$ &$10^4$ & 1.0 &0.14 &100.0&1\,000.0& 3\,367 $\pm$ 619 \\
35 & $100$ &$10^4$ & 1.0 &0.14 &100.0&5\,000.0& 3\,700 $\pm$ 820 \\

36 & $500$ &$10^4$ & 1.0 &0.14 &20.0&20.0& 683 $\pm$ 159 \\
37 & $500$ &$10^4$ & 1.0 &0.14 &20.0&50.0& 1\,180 $\pm$ 324 \\
38 & $500$ &$10^4$ & 1.0 &0.14 &20.0&100.0& 1\,653 $\pm$ 279 \\
39 & $500$ &$10^4$ & 1.0 &0.14 &20.0&200.0& 2\,287 $\pm$ 348 \\
40 & $500$ &$10^4$ & 1.0 &0.14 &20.0&500.0& 3\,310 $\pm$ 312 \\
41 & $500$ &$10^4$ & 1.0 &0.14 &20.0&1\,000.0& 3\,987 $\pm$ 208 \\
42 & $500$ &$10^4$ & 1.0 &0.14 &20.0&5\,000.0& 4\,027 $\pm$ 332 \\

43 & $1\,000$ &$10^4$ & 1.0 &0.14 &10.0&20.0& 645 $\pm$ 141 \\
44 & $1\,000$ &$10^4$ & 1.0 &0.14 &10.0&50.0& 1\,343 $\pm$ 215 \\
45 & $1\,000$ &$10^4$ & 1.0 &0.14 &10.0&100.0& 1\,785 $\pm$ 392 \\
46 & $1\,000$ &$10^4$ & 1.0 &0.14 &10.0&200.0& 2\,502 $\pm$ 166 \\
47 & $1\,000$ &$10^4$ & 1.0 &0.14 &10.0&500.0& 3\,622 $\pm$ 204 \\
48 & $1\,000$ &$10^4$ & 1.0 &0.14 &10.0&1\,000.0& 4\,092 $\pm$ 158 \\
49 & $1\,000$ &$10^4$ & 1.0 &0.14 &10.0&5\,000.0& 4\,317 $\pm$ 267 \\

50 & $5\,000$ &$10^4$ & 1.0 &0.14 &2.0&20.0& 181 $\pm$ 144 \tablefootmark{b} \\
51 & $5\,000$ &$10^4$ & 1.0 &0.14 &2.0&50.0& 552 $\pm$ 71 \\
52 & $5\,000$ &$10^4$ & 1.0 &0.14 &2.0&100.0& 1\,497 $\pm$ 189 \\
53 & $5\,000$ &$10^4$ & 1.0 &0.14 &2.0&200.0& 2\,264 $\pm$ 765 \\
54 & $5\,000$ &$10^4$ & 1.0 &0.14 &2.0&500.0& 3\,630 $\pm$ 166 \\
55 & $5\,000$ &$10^4$ & 1.0 &0.14 &2.0&1\,000.0& 4\,084 $\pm$ 213 \\
56 & $5\,000$ &$10^4$ & 1.0 &0.14 &2.0&5\,000.0& 5\,185 $\pm$ 236 \\
\hline
\end{tabular}
\tablefoot{Summary of simulations that include stellar mergers showing the number of particles, the cluster mass, the ratio of external potential to cluster mass, the virial radius, the initial stellar masses, the initial stellar radii and the average mass of the most massive star at the end of the simulation. 
\tablefoottext{a}{Value obtained as the average from 6 simulations.}
\tablefoottext{b}{Simulation run until 5\,657~$t_{\rm cross}=22$~Myr.}}
\end{table*}

\subsection{Stellar mergers}
\label{sec:stellar_mergers}
In our simulations, a merger occurs when two stars are separated by a distance equal to or smaller than the sum of their radii, this is:
\begin{equation}
    r\leq R_1+ R_2,
\end{equation}
with $r$ being the distance between both stars center of mass, and $R_1$ and $R_2$ are the radii of the stars. Given that the stars are physically in contact at this point, we also call this a stellar collision. Additionally, since each collision in our runs led to a merger, we then consider the terms merger and collision to be synonymous.
Once this condition was fulfilled, we merged both stars replacing them
by a new star whose total mass is the sum of the masses of the merging stars. Furthermore, the radius was calculated assuming that the star rapidly settles to a stable configuration where the new star has the same density as the previously merging stars, which is consistent with the calculations of \cite{Hosokawa2012} and \cite{Haemmerle2018}. Therefore the new mass and radius of the merger product were calculated as:
\begin{eqnarray}
        \label{eq:merger_new_mass}
        M_{\rm new} &=& M_1+M_2, \\
        \label{eq:merger_new_radius}
        R_{\rm new} &=& R_1\left( \frac{M_1 +M_2}{M_1}\right)^{1/3}.
\end{eqnarray}

\section{Results}
\label{sec:Results}
First we present the results of the simulations that do not include stellar mergers and describe the effects introduced by the addition of an external potential and the variation of the potential mass. Then, we present the results of our simulations that include stellar mergers and describe how their evolution and the formation of massive stars due to the runaway merger process are affected by the external potential.

\subsection{Simulations without mergers}
Here, we describe the main effects of including an external potential on the evolution of our star cluster models in the absence of stellar mergers. Due to the similarities between the runs, here, we only present the results for simulations number 5 and 8 listed in Table~\ref{tab:sim_no_mergers} with the setup described in Sec.~\ref{sec:sim_no_mergers}. However, a description for the remaining simulations listed in Table~\ref{tab:sim_no_mergers} is included in Appendix~\ref{sec:appendix2}. 

\begin{figure}
   \centering
   \includegraphics[width=\hsize]{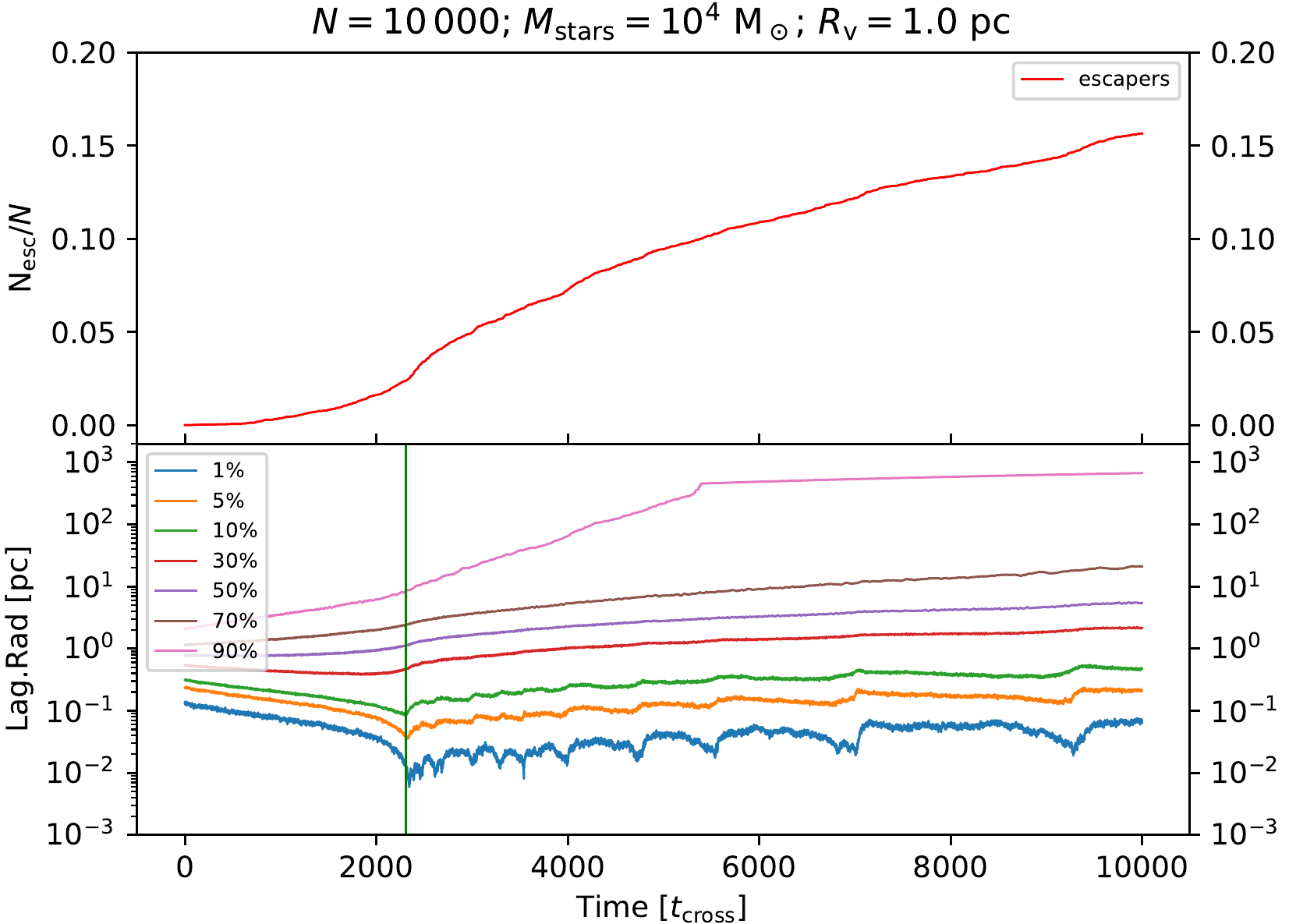}
      \caption{Evolution of a cluster with $N=10\,000$ stars with a total stellar mass of $M_{\rm stars}=10^4$~M$_\odot$ and virial radius $R_{\rm v}=1.0$~pc. The top panel shows the fraction of ejections from the cluster and the bottom panel shows the Lagrangian radii of the cluster. The vertical green line marks the moment of core collapse.}
         \label{fig:N10k_no_gas}
   \end{figure}

The evolution of clusters with $M_{\rm ext}=0 \ \times$~$M_{\rm stars}$ is well understood and consistent with the results of \cite{Spitzer1987}, where the clusters evolve toward core collapse as a result of two-body relaxation. The core then experiences core-oscillations, contracting and expanding again due to the formation of hard binaries that act as an energy source in the core. The core collapse is expected to occur between 10-20 half mass relaxation times $t_{\rm rh}$ that we calculated as \citep{Spitzer1987}:

\begin{equation}
\label{eq:tcc_spitzer}
  t_{\rm rh} = 0.138 \frac{N}{\ln{(\gamma N)}} t_{\rm cross},
\end{equation}
with $\gamma=0.4$ because we used equal mass stars and $t_{\rm cross}$ was calculated with Eq.(\ref{eq:tcross}) setting $M_{\rm tot}= M_{\rm stars}$. The half mass relaxation time for the cluster with $N=10\,000$ stars and $M_{\rm ext}=0\ \times$~$M_{\rm stars}$ is $t_{\rm rh}=166.38$~$t_{\rm cross}$ and core collapse occurs at 2\,306~$t_{\rm cross}$ which is 14~$t_{\rm rh}$. We found the core collapse time by visual inspection and looking for the first drop and subsequent rise in the 10\% Lagrangian radius as marked by the vertical green line in Fig.~\ref{fig:N10k_no_gas}.

At the moment of core collapse, the 10\% Lagrangian radius drops to 0.087~pc, yielding a mean density of 3.6$\times 10^5$~M$_{\odot}$~pc$^{-3}$. After this phase, the entire cluster begins to expand and the onset of ejections takes place with a total number of ejections of about 1\,500 until a time of 10\,000~$t_{\rm cross}$. A star is considered to be ejected from the cluster if its distance to the center of mass of the system is $>20$~$R_{\rm v}$ and its kinetic energy is higher than its potential energy at this location.

When we included an external potential in our simulations, we note first that, since the clusters start virialized, the velocities of the stars are higher compared to clusters without the background potential. This compensates for the extra gravitational force. Then we had to calculate the crossing time of the cluster using Eq.(\ref{eq:tcross}) with $M_{\rm tot} = M_{\rm stars} + M_{\rm ext}$.

   \begin{figure}
   \centering
   \includegraphics[width=\hsize]{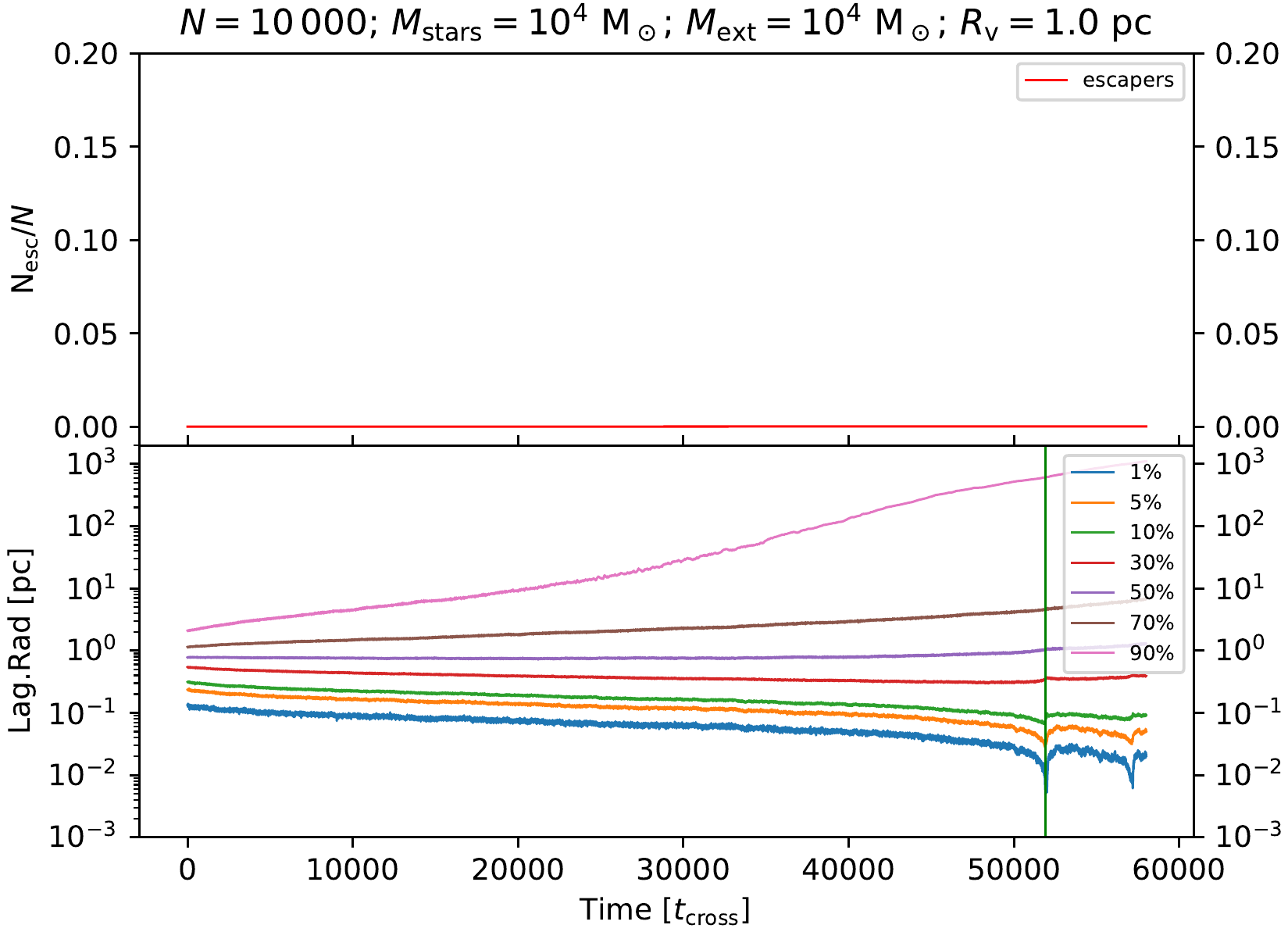}
      \caption{Evolution of a cluster with $N=10\,000$ stars with a total stellar mass of $M_{\rm stars}=10^4$~M$_\odot$ and a virial radius of $R_{\rm v}=1.0$~pc in the center of an external analytic potential with $M_{\rm ext}=10^4$~M$_\odot$. The top panel shows the fraction of ejections from the cluster and the bottom panel shows the Lagrangian radii of the cluster. The vertical green line marks the moment of core collapse.}
         \label{fig:N10k_Mg1}
   \end{figure}

We find that all of the clusters behave in a similar way. All of them evolve toward core collapse while there is little expansion of the outer parts (>30\% Lagrangian radii) compared to clusters without an external potential (see Fig.~\ref{fig:N10k_Mg1}).

For the case when the mass of the external potential is comparable to the total mass ($M_{\rm ext}=M_{\rm stars}$), the core collapse is delayed significantly, which occurs now at 51\,889~$t_{\rm cross}$ (see green vertical line in bottom panel of Fig.~\ref{fig:N10k_Mg1}). This is 23 times (in units of the cluster crossing time) later than the core collapse time for the cluster without a background potential. The 10\% Lagrangian radius at this moment drops to 0.073~pc, which yields a mean density of 6.0$\times 10^5$~M$_{\odot}$~pc$^{-3}$. For this simulation, the time is not long enough to see the contraction expansion phase that follows after core collapse. Finally, in this run, only one star is ejected from the cluster due to the higher escape velocity compared to the cluster without the external potential.

   \begin{figure}
   \centering
   \includegraphics[width=\hsize]{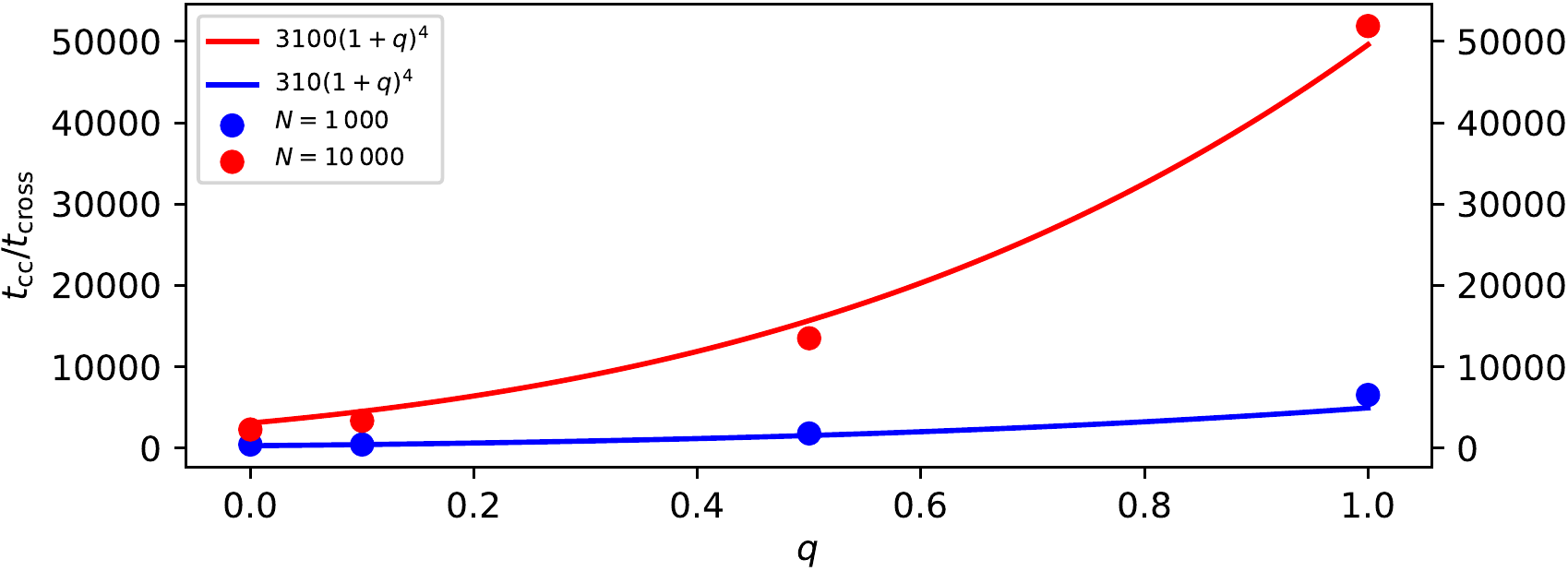}
      \caption{Core collapse time $t_{\rm cc}$ divided by the crossing time $t_{\rm cross}$ for clusters with $N=1\,000$ (blue circles) and $N=10\,000$ (red circles) as a function of $q=M_{\rm ext}/M_{\rm stars}$. We also plotted two relations that scale as $(1+q)^4$ which fit our data and therefore suggest that if the core collapse time is proportional to the relaxation time of the cluster, then the relaxation time scales as $(1+q)^4$, which is the same relation that we derive in Appendix~\ref{sec:appendix1}.}
         \label{fig:tcc_tcross}
   \end{figure}

In order to understand the delay in the core collapse time when including a background potential, we define the parameter $q=M_{\rm ext}/M_{\rm stars}$ as the ratio between the mass of the external potential and the total mass in stars, then we find that the core collapse time scales as $t_{\rm cc} \propto (1+q)^4$~$t_{\rm cross}$ as presented in Fig.~\ref{fig:tcc_tcross}. If we assume, as found by \cite{Spitzer1987} that the core collapse time is proportional to the relaxation time, then according to our data the relaxation time is proportional to $t_{\rm rh} \propto (1+q)^4$, which is consistent with Eq.(\ref{eq:trelax_on_q_tcros}) for which we derived the relaxation time of a cluster in the center of an analytic external potential (see Appendix~\ref{sec:appendix1}):
\begin{equation}
\label{eq:trelax_q}
    t_{\rm rh} = 0.138 \frac{N(1+q)^4}{\log{(\gamma N)}}  t_{\rm cross},
\end{equation}
where the crossing time must be calculated from Eq.(\ref{eq:tcross}) with $q=M_{\rm ext}/M_{\rm stars}$ (see the Appendix in \cite{Leigh2013} and \cite{Leigh2014} for similar derivations but slightly modified to consider different cases).

\begin{figure}
   \centering
   \includegraphics[width=\hsize]{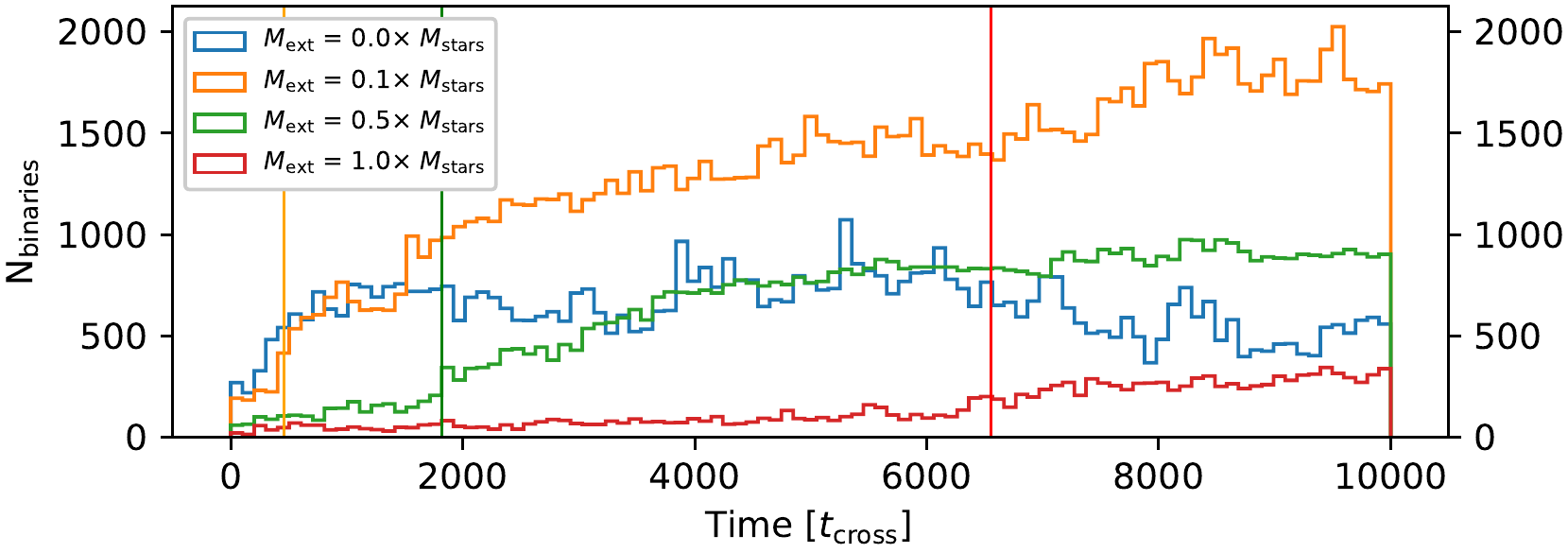}
      \caption{Number of binary systems as a function of time and for different values for the mass of the external potential $M_{\rm ext}$ for simulations with $N=1\,000$ and up to 8\,000 crossing times. The vertical lines mark the moments of core collapse and coincides with the onset of the formation of binary systems.}
         \label{fig:Nbin_time}
   \end{figure}

\begin{figure}
   \centering
   \includegraphics[width=\hsize]{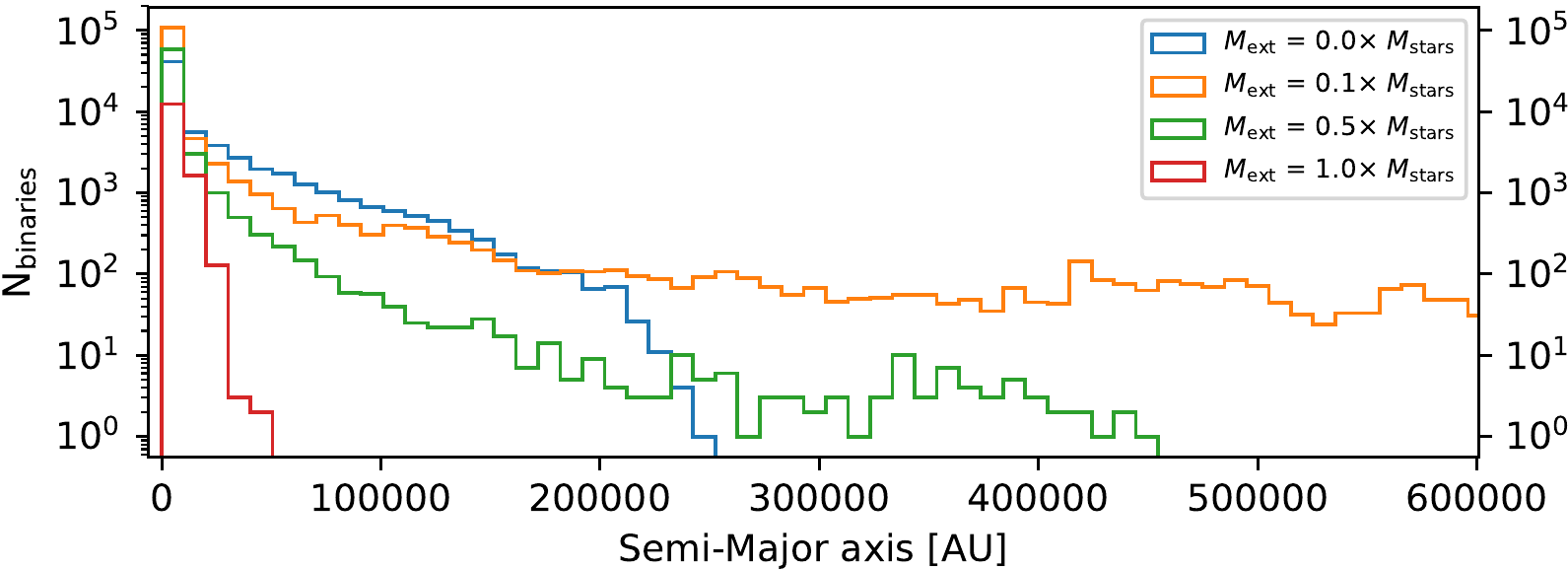}
      \caption{Distribution of the semi-major axis of the binary population for the set of simulations listed in Table~\ref{tab:sim_no_mergers}. The data clearly shows that increasing the mass of the external potential the formed binaries tend to be more tightly bound. Despite the fact that a low mass of the external potential ($M_{\rm ext}<1.0 \times$~$M_{\rm stars}$) may promote the formation of wider binaries, these systems tend to be short-lived due to the higher velocities of all the stars. This figure is only intended to show the distribution of the semi-major axis and not the total number of binaries that formed.}
         \label{fig:Nbin_semiaxis}
   \end{figure}

A second important effect that we note is on the binary population. Our results show that in increasing the mass of the background potential, the number of binary systems that form decreases dramatically as shown in Fig.~\ref{fig:Nbin_time} due to the increased mean kinetic energy of the stars. Consequently, we also find that these binaries tend to be more tightly bound (see Fig.~\ref{fig:Nbin_semiaxis}) given that they form at the hard-soft boundary, which is smaller for higher velocity dispersion \citep{Leigh2015}. This will certainly have an impact on the growth of a massive star through stellar mergers given that collisions are more likely to occur in binary systems.

\subsection{Simulations that include stellar mergers}
In the following, we present the results that we obtained when we included stellar mergers in the evolution of star clusters with and without an external Plummer potential. Additionally, we describe the effects introduced by this background potential on the formation of massive merger products. These simulations are listed in Table~\ref{tab:sim_mergers}.

The first effect that we note is a delay in the runaway growth of the central object, which is explained by the delay in the core collapse time due to a larger relaxation time when increasing the mass of the potential, as can be seen in Eq.(\ref{eq:trelax_q}). Moreover with this external force, binary systems are harder to form, as shown in Fig.~\ref{fig:Nbin_time} and these binaries tend to be more compact.

   \begin{figure}
   \centering
   \includegraphics[width=\hsize]{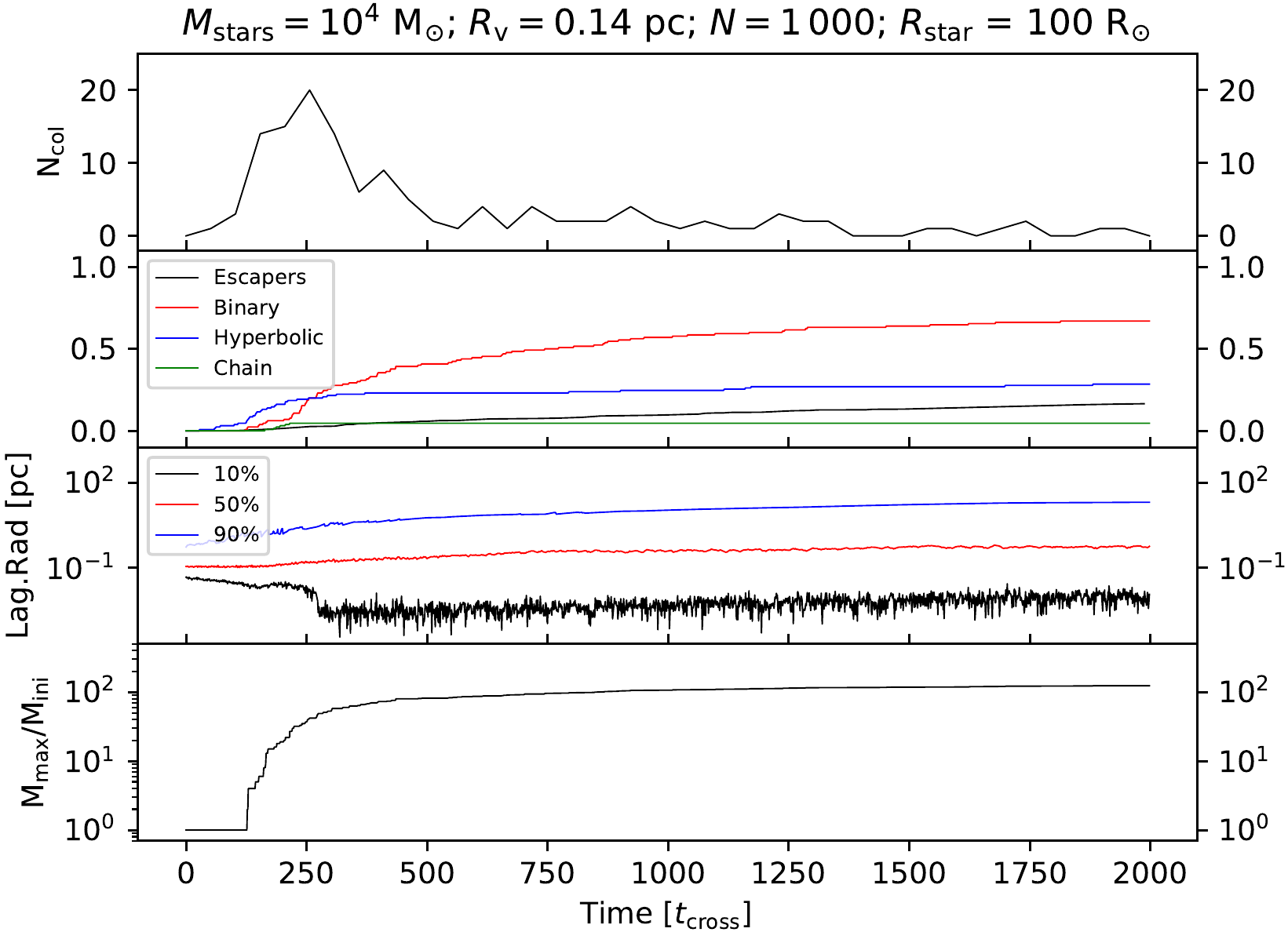}
      \caption{Evolution of a cluster with $N=1\,000$ stars, total mass $M_{\rm stars}=10^4$~M$_\odot$, $R_{\rm v}=0.14$ pc, and $R_{\rm star}=100$~R$_\odot$. In the uppermost panel, we present the number of mergers as a function of time, the fraction of binary, hyperbolic, and chain mergers in the second panel along with the fraction of stars that escape from the cluster. The third panel shows the 10,50 and 90\% Lagrangian radii, and the bottom panel illustrates the mass growth of the most massive object that formed through mergers in the cluster.}
         \label{fig:N1k_Mg0_R100_merger}
   \end{figure}

\begin{figure}
   \centering
   \includegraphics[width=\hsize]{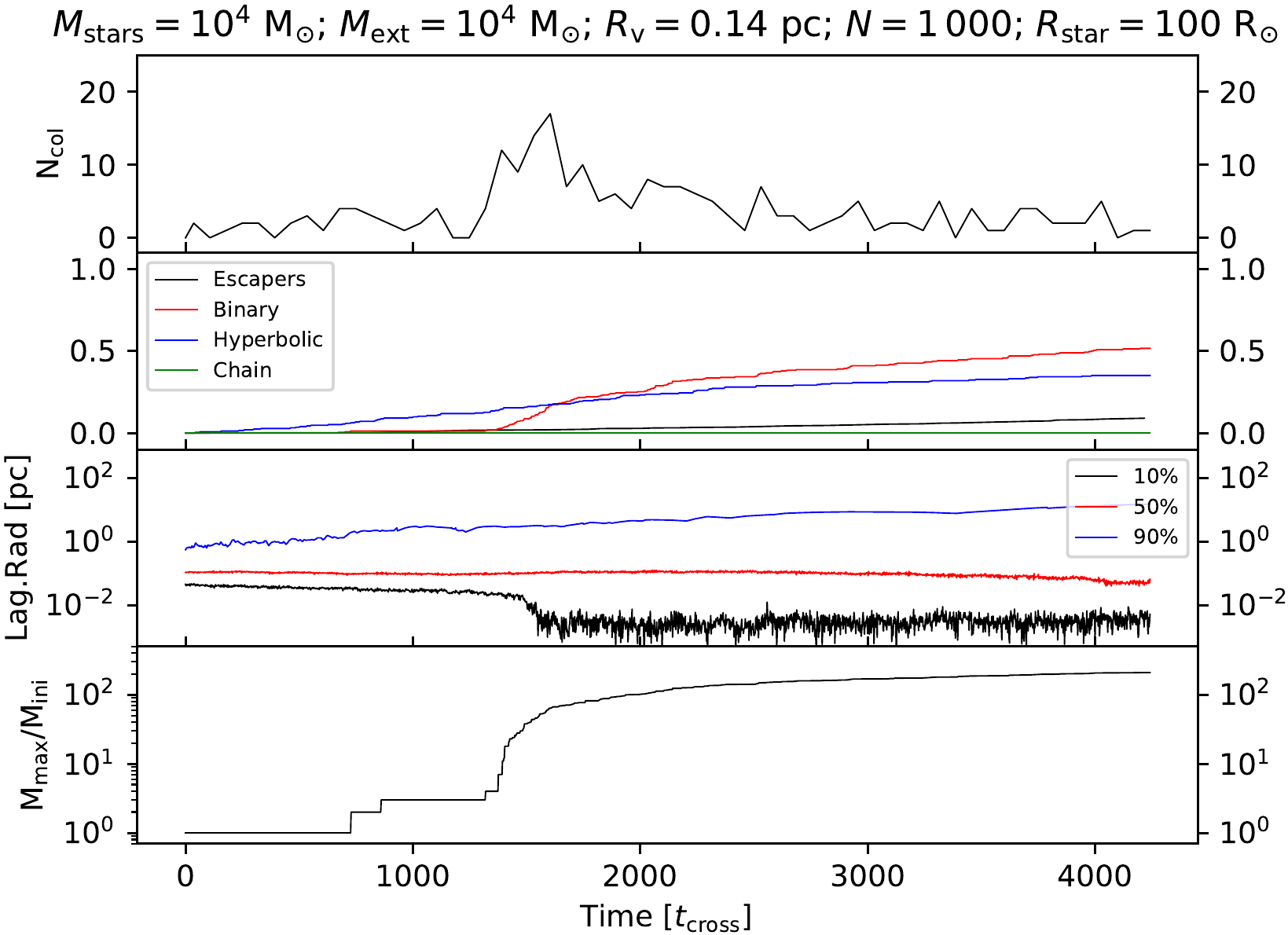}
      \caption{Evolution of a cluster with $N=1\,000$ stars, stellar mass $M_{\rm stars}=10^4$~M$_\odot$, $R_{\rm v}=0.14$ pc, and $R_{\rm star}=100$~R$_\odot$ in the center of an external potential with $M_{\rm ext}=10^4$~M$_\odot$. In the uppermost panel, we present the number of mergers as a function of time, the fraction of binary, hyperbolic, and chain mergers in the second panel, along with the fraction of stars that escape from the cluster. The third panel shows the 10, 50, and 90\% Lagrangian radii, and the bottom panel illustrates the mass growth of the most massive object that formed through mergers in the cluster. The drop in the 50\% Lagrangian radius toward the end of the simulation suggests that the central object may grow up to 5\,000~M$_\odot$.}
         \label{fig:N1k_Mg1_R100_merger}
   \end{figure}

Here we describe the time evolution for two clusters with $N=1\,000$ stars, with and without an external potential, which correspond to simulations number 17 and 45 in Table~\ref{tab:sim_mergers} and we also present their evolution in Figs.~\ref{fig:N1k_Mg0_R100_merger} and \ref{fig:N1k_Mg1_R100_merger}, respectively. These simulations serve as examples of the typical evolution of the runs that include stellar mergers, given that all the simulated clusters follow very similar patterns. \\
In our runs, we identify three types of mergers: namely hyperbolic mergers, which are mergers that occur between stars that are not gravitationally bound; binary mergers, which occur between stars that are part of a binary system; and chain mergers, which occur between stars that are part of a higher order system, for example, triples.
Before the onset of runaway mergers occurs in all of the clusters, there is a nearly constant merger rate for hyperbolic mergers which do not produce a single massive star, but instead several less massive stars that eventually sink to the cluster center, thus finally contributing to the growth of the most massive star. We note that the hyperbolic merger rate in clusters that do not include an external potential is only maintained for short periods of time (see Fig.~\ref{fig:N1k_Mg0_R100_merger}); whereas in clusters that include the extra force, the hyperbolic merger rate is maintained until the end of the simulation (see Fig.~\ref{fig:N1k_Mg1_R100_merger}). Chain mergers are very rare with only a handful having been identified among all of our simulations.

During the period before the rapid growth of the most massive star, relaxation processes drive an energy flow from the central parts of the cluster (inside the half-mass radius) via high velocity stars that migrate to the cluster halo and cause the expansion of the outer parts. All of the clusters evolve toward core collapse, but not all of them reach that stage because for large stellar radii $R_{\rm star} \geq 100$~R$_\odot$, the growth of the central star is fast and the drop in the 10\% Lagrangian radius that we see in Fig.~\ref{fig:N1k_Mg0_R100_merger} occurs because the central object has already gathered 10\% of the cluster mass, and thus we see the position of the central star rather than a collapsed core. The onset of growth for the central star is also marked by a rapid increase in the merger rate of stars in binary systems despite the presence of an external potential. 

An important difference that we note when including the background potential is on the expansion of the clusters and the fraction of escaping stars. In this sense we found that the external potential prevents the evaporation of the cluster (see panel 2 of Fig.~\ref{fig:N1k_Mg1_R100_merger}) and keeps it compact so the mergers can continue for longer periods of time, potentially gathering up to 50\% or more of the cluster mass in the central object, as the third panel of Fig.~\ref{fig:N1k_Mg1_R100_merger} suggests.

\subsection{Formation and growth of the most massive star}
One important goal of this research is to address the effect of the external potential on the formation of a massive star by stellar mergers. Thus, we present the results regarding the mass growth of the most massive star in our clusters, along with a description of the effects introduced by the background potential.

The first important effect is a delay in the formation of the most massive star due to the increased stellar velocity, which in turn causes relaxation processes to be slower. This then leads to a reduction in the number of binary systems that form, at a given time, if the mass of the external potential is comparable to the total stellar mass.
On the other hand, the external potential can also favor the mass growth of the central star by preventing the evaporation and expansion of the cluster. 

In the following, we present a model that we used to estimate the mass of the central object at different times for both a cluster with and without an external potential. For this, we followed the same method described in \cite{Reinoso2018}.

   \begin{figure}
   \centering
   \includegraphics[width=\hsize]{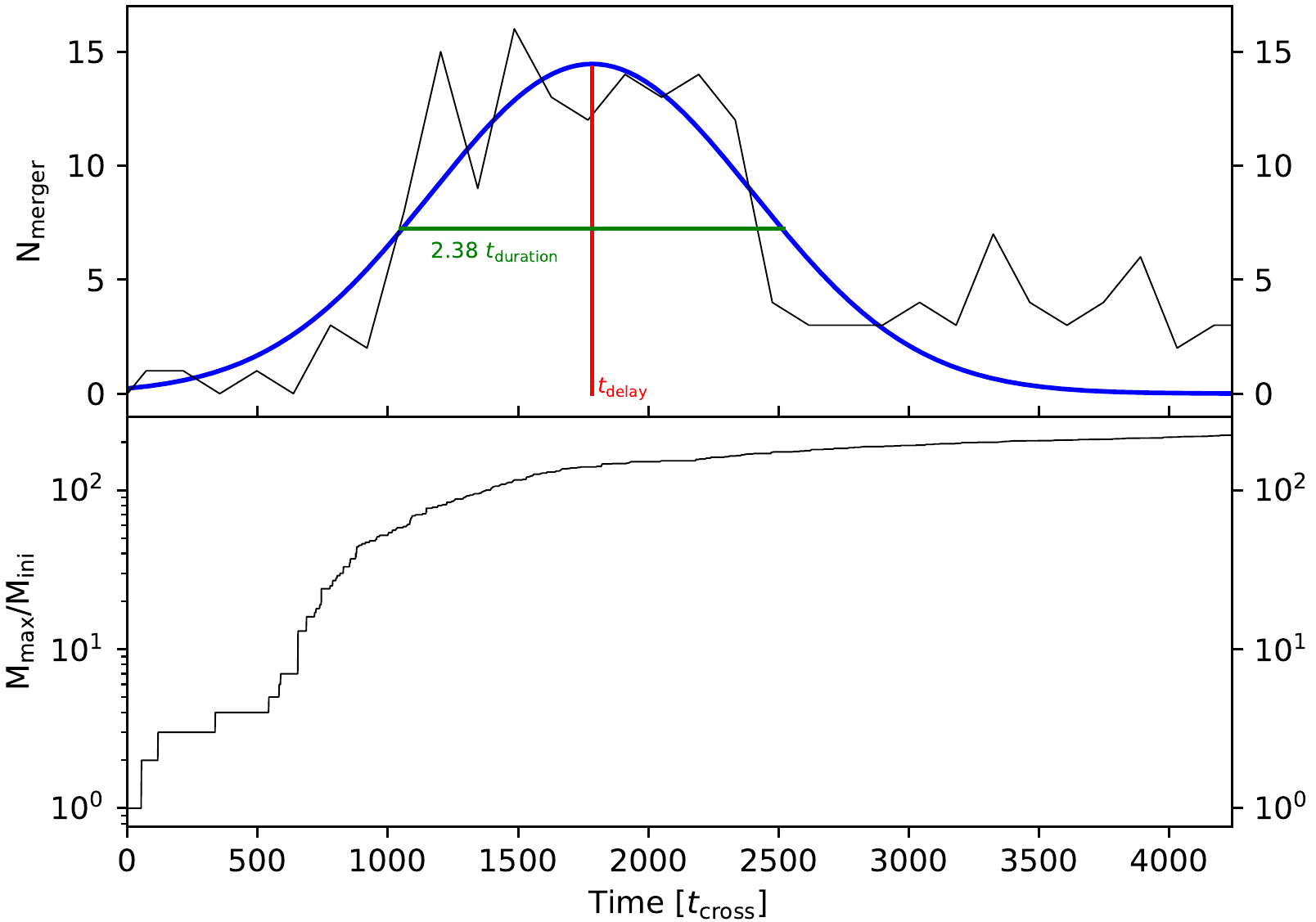}
      \caption{Example of a Gaussian fit to the number of mergers with the central star for simulation number 45 listed in Table~\ref{tab:sim_mergers}. In the top panel, we show the number of mergers with the most massive star as a function of time (black solid line) along with the Gaussian function that we used to estimate the total number of mergers at different times. The parameters $t_{\rm delay}$ and $t_{\rm duration}$ are also shown in the figure. The bottom panel shows the mass growth of the central star whose rapid growth coincides with the peak in the number of mergers.}
         \label{fig:Gausian_example}
   \end{figure}

In the lower panel of Fig.~\ref{fig:Gausian_example}, we first present an example of the mass growth of the most massive star, which undergoes a very rapid growth at around 1\,000-1\,500~$t_{\rm cross}$. This coincides with the peak in the number of mergers with the central object as shown in the top panel of the same figure.

\begin{figure}
   \centering
   \includegraphics[width=\hsize]{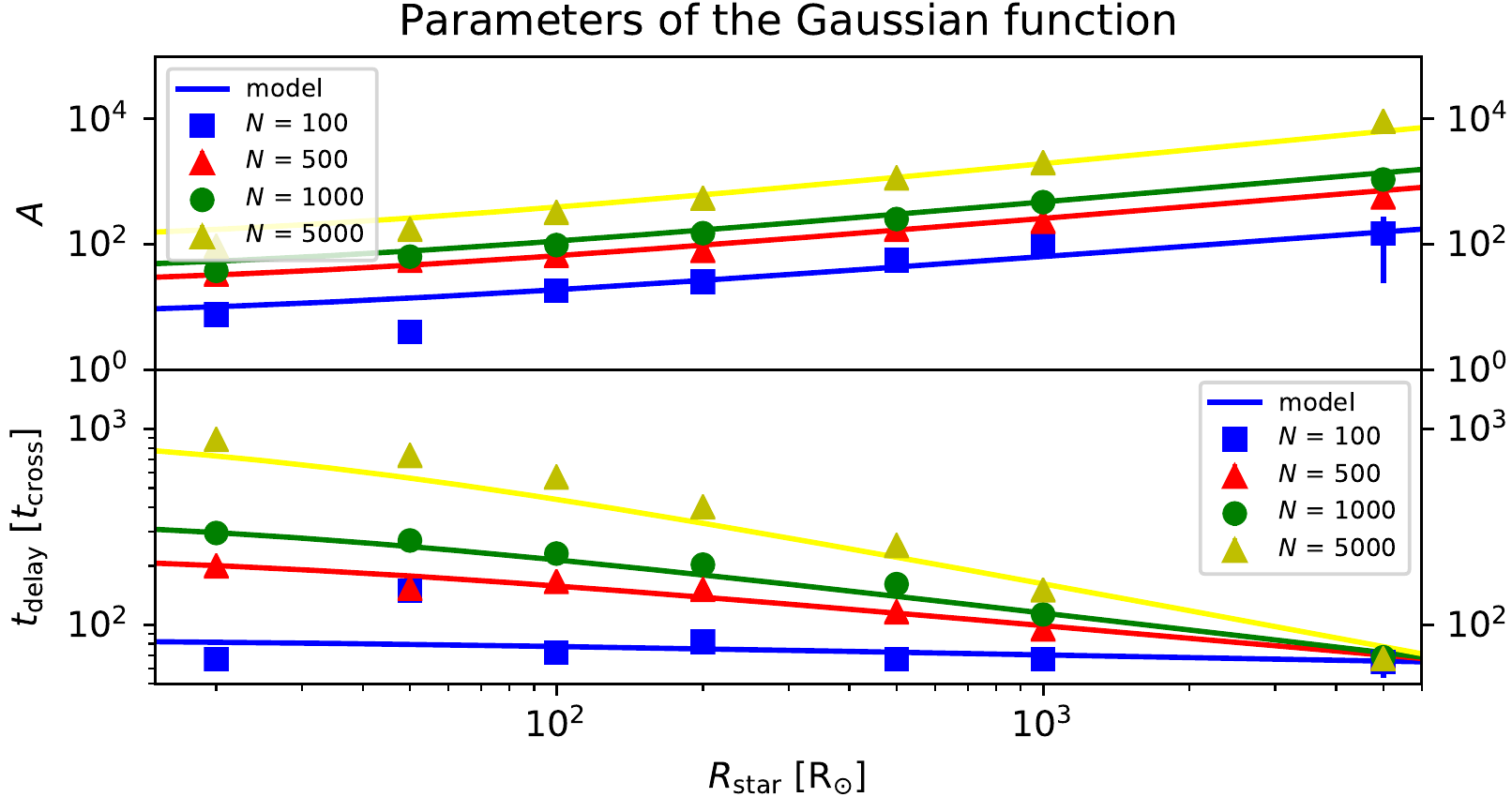}
      \caption{Parameters of the Gaussian fit to the number of mergers with the central star for simulations of clusters without an external potential. The data points were obtained after the fit to the combined data of six simulations. In the upper panel, we show the values of the normalization factor $A$ for different values of $N$ and $R_{\rm star}$ along with our model from Eq.(\ref{eq:A_no_gas}). These data points show that the number of mergers increases with both $N$ and $R_{\rm star}$ as expected. The bottom panel shows the different values of $t_{\rm delay}$ in addition to the different values of $N$ and $R_{\rm star}$ along with the fitting function presented in Eq.(\ref{eq:tdelay_no_gas}). These points show that $t_{\rm delay}$ decreases with $R_{\rm star}$, but it increases with $N$.}
         \label{fig:parameters_no_gas}
   \end{figure}

\begin{figure}
   \centering
   \includegraphics[width=\hsize]{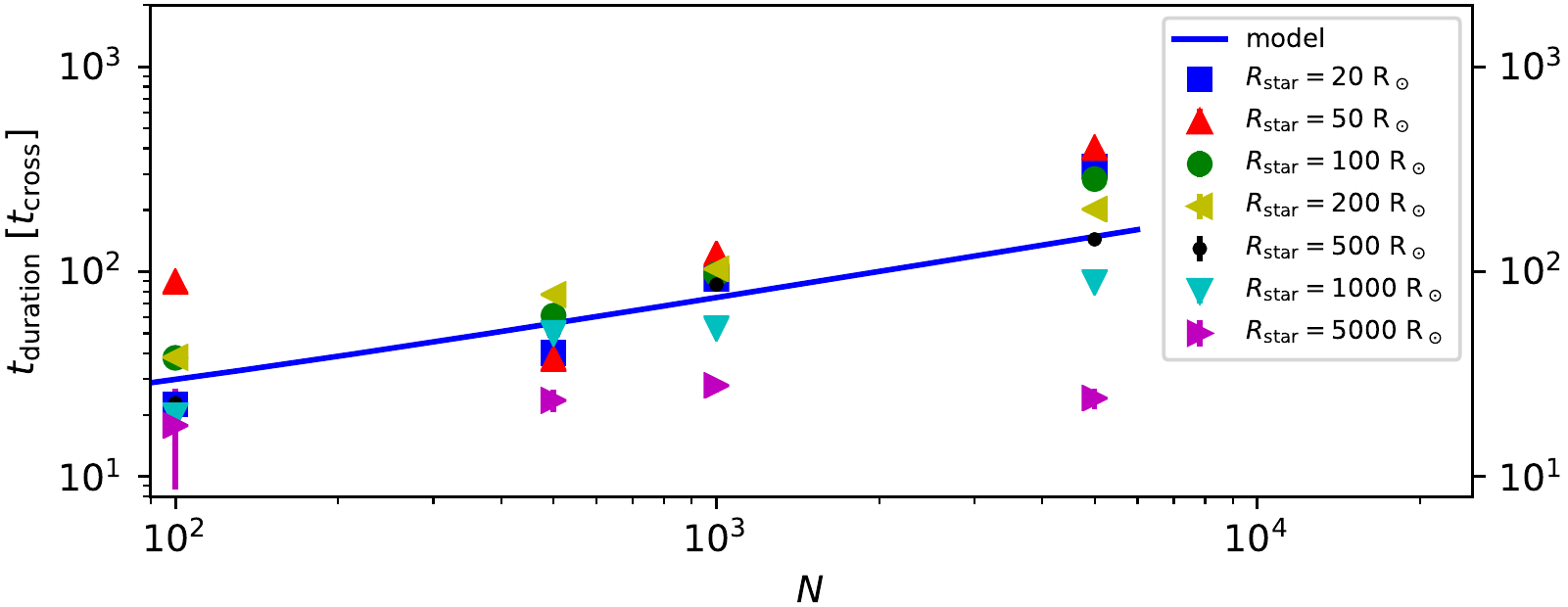}
      \caption{$t_{\rm duration}$ as a function of the number of stars $N$ for clusters without an external potential. We found no clear correlations between $t_{\rm duration}$ and $R_{\rm star}$. The solid blue line shows the fitting function from Eq.(\ref{eq:t_duration_no_gas}).}
         \label{fig:tduration_vs_N}
   \end{figure}

\begin{figure}
   \centering
   \includegraphics[width=\hsize]{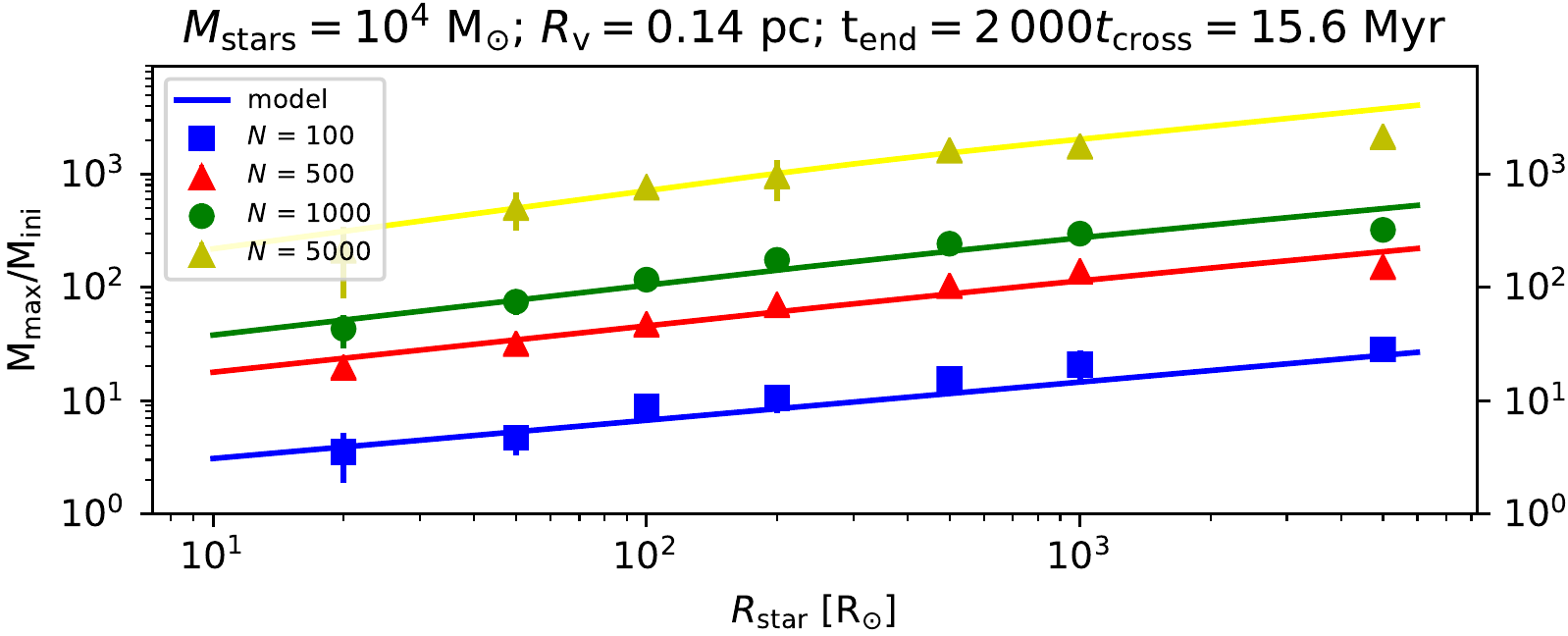}
      \caption{Mean mass of the most massive star normalized by its initial mass, as a function of $N$ and $R_{\rm star}$ for clusters without an external potential evolved until $2\,000$~$t_{\rm cross}=15.6$~Myr. The symbols represent the mean values obtained from six simulations and the solid lines show the results of the calculations with our model presented in Eq.(\ref{eq:Nmerger_total}) using $t=15.6$~Myr and the parameters obtained from Eqs.(\ref{eq:A_no_gas}), (\ref{eq:tdelay_no_gas}), (\ref{eq:t_duration_no_gas}), and (\ref{eq:D_no_gas}).
      }
         \label{fig:Mmax_no_gas}
   \end{figure}

\begin{figure}
   \centering
   \includegraphics[width=\hsize]{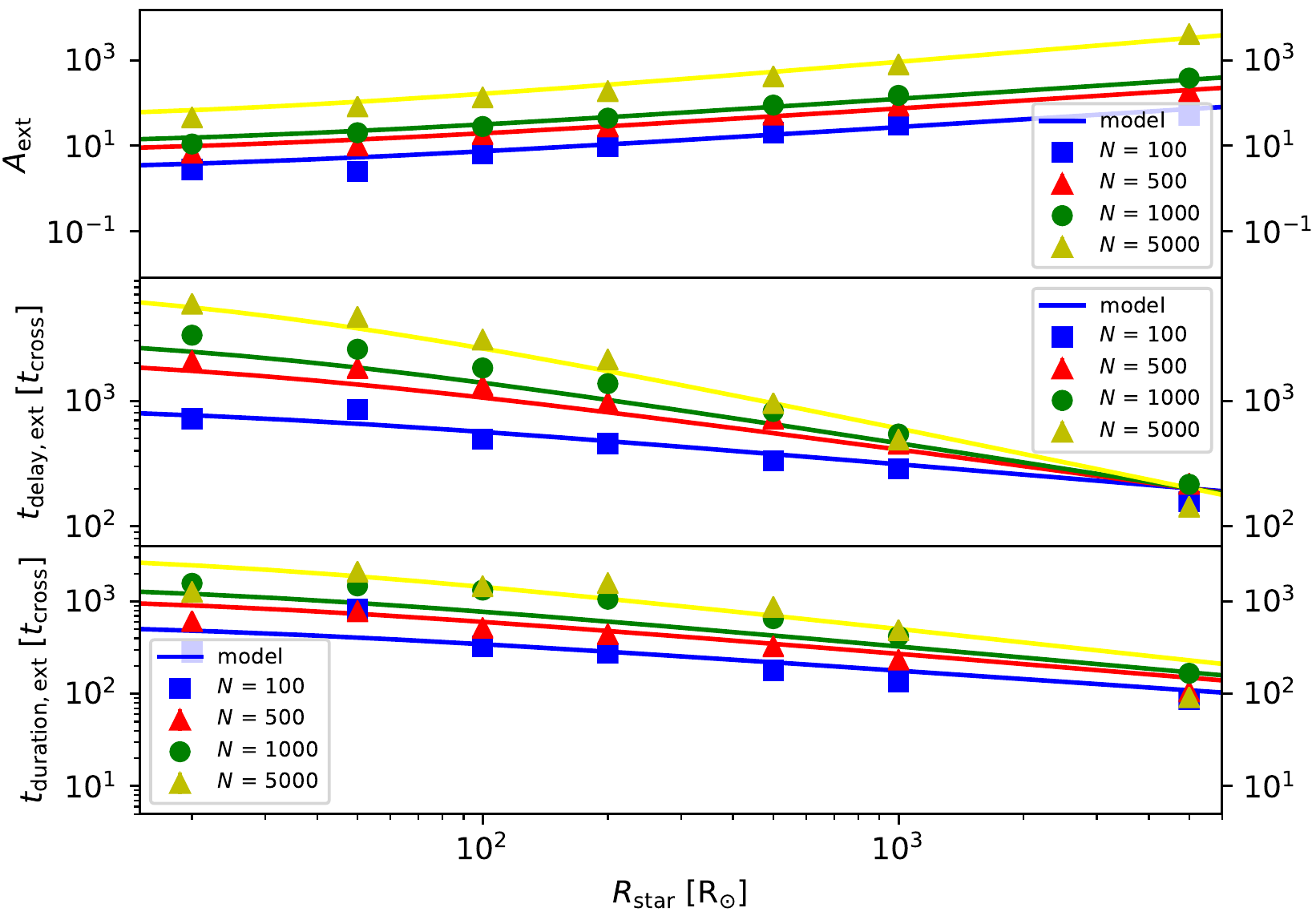}
      \caption{Parameters of the Gaussian fit to the number of mergers with the central star for simulations of clusters with an external potential. The data points were obtained after the fit to the combined data of six simulations. In the upper panel, we show the values of the normalization factor $A_{\rm ext}$ for different values of $N$ and $R_{\rm star}$ along with our model from Eq.(\ref{eq:Nmerger_total}). These data points show that the number of mergers increases with both $N$ and $R_{\rm star}$ as expected. The middle panel shows the different values of $t_{\rm delay,ext}$ in addition to the different values of $N$ and $R_{\rm star}$ along with the fitting function presented in Eq.(\ref{eq:tdelay_gas}). These points show that $t_{\rm delay,ext}$ decreases with $R_{\rm star}$, but it increases with $N$. Additionally, these values are larger than the values of $t_{\rm delay}$ for clusters without an external potential (see Fig.~\ref{fig:parameters_no_gas}). The lower panel shows the values of $t_{\rm duration,ext}$ along with the fitting function from Eq.(\ref{eq:t_duration_gas}). We clearly see a positive correlation with $N$ and a negative correlation with $R_{\rm star}$.}
         \label{fig:parameters_gas}
   \end{figure}

\begin{figure}
   \centering
   \includegraphics[width=\hsize]{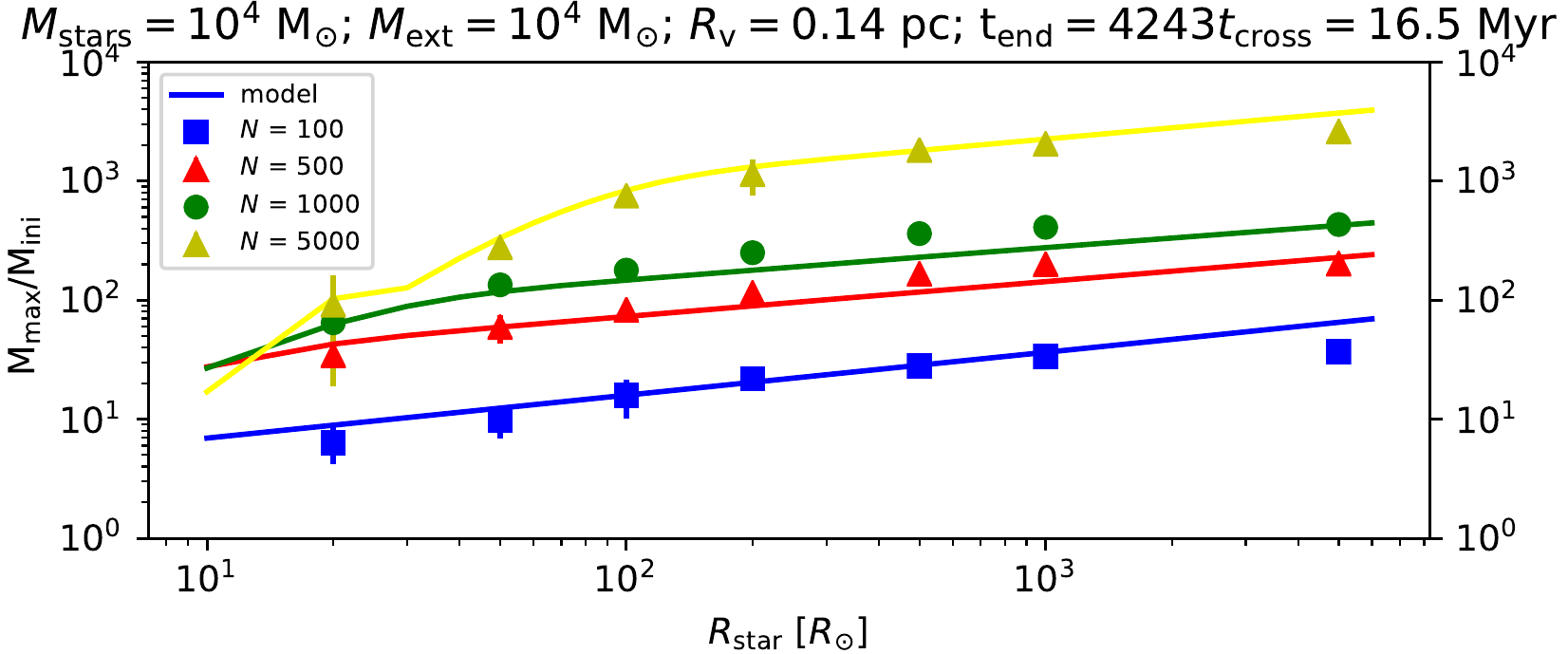}
      \caption{Mean mass of the most massive star normalized by its initial mass, as a function of N and $R_{\rm star}$ for clusters without an external potential evolved until $3\,000$~$t_{\rm cross}=16.5$~Myr, except for the point at $N=5\,000$ and $R_{\rm star}=20$~R$_\odot$, which are simulations that were evolved until $4\,000$~$t_{\rm cross}=22$~Myr. The symbols represent the mean values obtained from six simulations and the solid lines show the results of the calculations with our model presented in Eq.(\ref{eq:Nmerger_total}) using $t=16.5$~Myr ($t=22$~Myr for $N=5\,000$ and $R_{\rm star}<50$~R$_\odot$) and the parameters obtained from Eqs.(\ref{eq:A_gas}), (\ref{eq:tdelay_gas}), (\ref{eq:t_duration_gas}) and (\ref{eq:D_gas}).
      }
         \label{fig:Mmax_gas}
   \end{figure}

We fit the Gaussian function presented in Eq.(\ref{eq:gausianfit}) to the combined data from the six random realizations for each simulation setup presented in Table~\ref{tab:sim_mergers} in order to get an estimate for the number of mergers with the central object during the rapid growth. Thus, we define the delay time $t_{\rm delay}$ as the time at which the peak in the Gaussian occurs, and the height $A$ of the Gaussian gives us an estimate for the number of mergers at $t_{\rm delay}$. Additionally, we define the duration time $t_{\rm duration}=FWHM/\left(2\sqrt{2\ln{2}}\right)$, with $FWHM$ being the full width at half maximum of the Gaussian. An illustration of the fitting formula and two parameters are shown in the top panel of Fig.~\ref{fig:Gausian_example}. By doing this, we were able to obtain an estimate for the moment at which there is rapid growth of the most massive star ($t_{\rm delay}$) and an estimate for the duration of this period ($t_{\rm duration}$), which we can later combine with $A$ and a normalization factor $D$ to finally obtain the number of mergers experienced by the central star using Eq.(\ref{eq:Nmerger_total}) and therefore an estimate for its final mass at different times. Furthermore we determined $D$ by comparing our model to the results from our simulations. We also show how well our model reproduces the data of our simulations in Figs.~\ref{fig:Mmax_no_gas} and \ref{fig:Mmax_gas}. The total number of mergers at time $t$ can thus be calculated as:
\begin{eqnarray}
\label{eq:gausianfit}
N_{\rm merger}(t)&=&A \exp{\left(-\frac{(t-t_{\rm delay})^2}{2t_{\rm duration}^2} \right)},
\end{eqnarray}
and the total number of mergers until a time $t_{\rm end}$ can be calculated as:
\begin{eqnarray}
\label{eq:Nmerger_total}
N_{\rm merger,total} &=&AD\displaystyle \int_{0}^{t_{\rm end}}  \exp{\left( -\frac{(t-t_{\rm delay})^2)}{2t_{\rm duration}^2}\right)} dt.
\end{eqnarray}
Although the data is not a perfect Gaussian (see Fig.~\ref{fig:Gausian_example}), we only need an estimate for the delay time $t_{\rm delay}$, the duration time $t_{\rm duration}$, and an estimate for the number of collisions $A$ at $t_{\rm delay}$. Following this procedure, we first combined the data of the six random realizations per each of the configurations listed in Table~\ref{tab:sim_mergers} and then applied the Gaussian fit of Eq.(\ref{eq:gausianfit}) to find the values of $A$, $t_{\rm delay}$, and $t_{\rm duration}$ that we present in Figs.~\ref{fig:parameters_no_gas} and \ref{fig:tduration_vs_N} for clusters with $M_{\rm ext}/M_{\rm stars} =0.0$ (simulations 1-28 in Table~\ref{tab:sim_mergers}) and in Fig.~\ref{fig:parameters_gas} for clusters with $M_{\rm ext}/M_{\rm stars} =1.0$ (simulations 29-56 in Table~\ref{tab:sim_mergers}). Then we fit these data using an implementation of the nonlinear least-squares Marquardt-Levenberg algorithm in \texttt{gnuplot} and obtained Eqs.(\ref{eq:A_no_gas}), (\ref{eq:tdelay_no_gas}), and (\ref{eq:t_duration_no_gas}) for clusters without an external potential and Eqs.(\ref{eq:A_gas}), (\ref{eq:tdelay_gas}) and (\ref{eq:t_duration_gas}) for clusters in a background potential.
Using these equations, we were able to compute $N_{\rm merger,total}$ assuming first $D=1$ and we compared the calculated values to the real values from the simulations to adjust $D$ and reproduce the results correctly. By doing this, we obtain Eqs.(\ref{eq:D_no_gas}) and (\ref{eq:D_gas}):

\begin{eqnarray}
\label{eq:A_no_gas}
\log_{10}{(A)} &=&[0.104\log_{10}{({R_{\rm star}})}+0.562]\log_{10}{(N)} \nonumber \\
 & &+0.358\log_{10}{({R_{\rm star}})}-1.011, \\
 \label{eq:tdelay_no_gas}
 \log_{10}{(t_{\rm delay}[t_{\rm cross}])}&=&[-0.246\log_{10}{(N)}+0.443]\log_{10}{({R_{\rm star}})} \nonumber \\
 & &+0.954\log_{10}{(N)}+0.077, \\
 \label{eq:t_duration_no_gas}
 \log_{10}{(t_{\rm duration}[t_{\rm cross}])}&=& 0.430\log_{10}{(N)}+0.580,\\
 \label{eq:D_no_gas}
 \log_{10}{(D)}&=&-0.227\log_{10}{({R_{\rm star}})}-1.791,
\end{eqnarray}
and the parameters for clusters in an external potential:
\begin{eqnarray}
\label{eq:A_gas}
\log_{10}{(A_{\rm ext})} &=&[3.975\times10^{-5}N+0.606]\log_{10}{({R_{\rm star}})} \nonumber \\
 & &+0.554\log_{10}(N)-1.510,  \\
 \label{eq:tdelay_gas}
 \log_{10}{(t_{\rm delay,ext}[t_{\rm cross,ext}])}&=&[-0.006\log_{10}(N)+0.240]\log_{10}({R_{\rm star}}) \nonumber \\
 & &+0.064\log_{10}{(N)}-1.170 +t_{\rm delay}, \\
 \label{eq:t_duration_gas}
 \log_{10}{(t_{\rm duration,ext}[t_{\rm cross,ext}])}&=&-0.178 N^{0.119}\log_{10}{({R_{\rm star}})} \nonumber \\
 & &+2.163 N^{0.073}+0.151, \\
 \label{eq:D_gas}
 \log_{10}{(D_{\rm ext})}&=&0.053\log_{10}{({R_{\rm star}})} \nonumber \\ 
 & & -1.879\times10^{6.9N/10^6} -0.778.
\end{eqnarray}

It is important to note that in Eqs.(\ref{eq:tdelay_no_gas}) and (\ref{eq:t_duration_no_gas}), the delay time t$_{\rm delay}$ and duration time t$_{\rm duration}$ are expressed in units of the crossing time of the cluster defined in Eq.(\ref{eq:tcross}) and consequently when calculating the total number of mergers until a time t$_{\rm end}$ using Eq.(\ref{eq:Nmerger_total}), this time must also be expressed in units of the crossing time $t_{\rm cross}$. The same principle applies for Eqs.(\ref{eq:tdelay_gas}) and (\ref{eq:t_duration_gas}). That is, when calculating the number of mergers until a time t$_{\rm end}$ using Eq.(\ref{eq:Nmerger_total}), this time must be expressed in units of the cluster crossing time that have been modified by the mass of the external potential $t_{\rm cross,ext}$, which is defined in Eq.(\ref{eq:tcross}).

The delay time for clusters in a background potential is defined in Eq.(\ref{eq:tdelay_gas}) as $t_{\rm delay,ext}=t_{\rm delay}+t_{{\rm delay},M_{\rm ext}=M_{\rm stars}}$, which is the delay time for clusters without an external potential $t_{\rm delay}$ plus an additional term $t_{{\rm delay},M_{\rm ext}=M_{\rm stars}}=[-0.006\log_{10}(N)+0.240]\log_{10}({R_{\rm star}})+0.064\log_{10}{(N)}-1.170$ that, in principle, may depend on the mass of the potential. However we do not examine this potential dependence here.

To help in the reading of the equations, we define a general equation of the form:

\begin{equation}
 \label{eq:parameters_general}
    \log_{10}(x)=\alpha \log_{10}(R_{\rm star})+ \beta \log_{10}(N)+ \gamma
\end{equation}
and present the parameters $\alpha$, $\beta$, and $\gamma$ in Table~\ref{tab:parameters_equations}. They can be used to recover Eqs.(\ref{eq:A_no_gas})-(\ref{eq:D_gas}).

\begin{table*}
\caption{Parameters for Eq.(\ref{eq:parameters_general})} 
\label{tab:parameters_equations}
\centering
\begin{tabular}{rrrr}
\hline \hline
$x$ & $\alpha$ & $\beta$& $\gamma$\\
\hline
$A$                & 0.358 & $0.140\log_{10}(R_{\rm star})+0.562$  &-1.011\\
$t_{\rm delay}[t_{\rm cross}]$    & $-0.246\log_{10}(N)+0.443$ &0.954 & 0.077\\
$t_{\rm duration}[t_{\rm cross}]$ & 0 & 0.430 & 0.580\\
$D$                & -0.227 & 0 &-1.791 \\

$A_{\rm ext}$          & $3.975\times 10^{-5}N +0.606$ &0.554 & -1.510\\
$t_{\rm delay,ext}[t_{\rm cross,ext}]$    & $-0.006\log_{10}(N)+0.240$ &0.064 & -1.170+$t_{\rm delay}[t_{\rm cross}]$\\
$t_{\rm duration,ext}[t_{\rm cross,ext}]$ & $-0.178N^{0.119}$ &0 & $2.163N^{0.073}$+0.151\\
$D_{\rm ext}$          & $0.053$ &0 & $-1.879\times 10^{6.9N/10^6} -0.778$\\
\hline
\end{tabular}
\tablefoot{Parameters for Eq.(\ref{eq:parameters_general}) that can be used to obtain Eq.'s(\ref{eq:A_no_gas})-(\ref{eq:D_gas}).}
\end{table*}

\begin{figure*}
   \centering
   \includegraphics[width=\hsize]{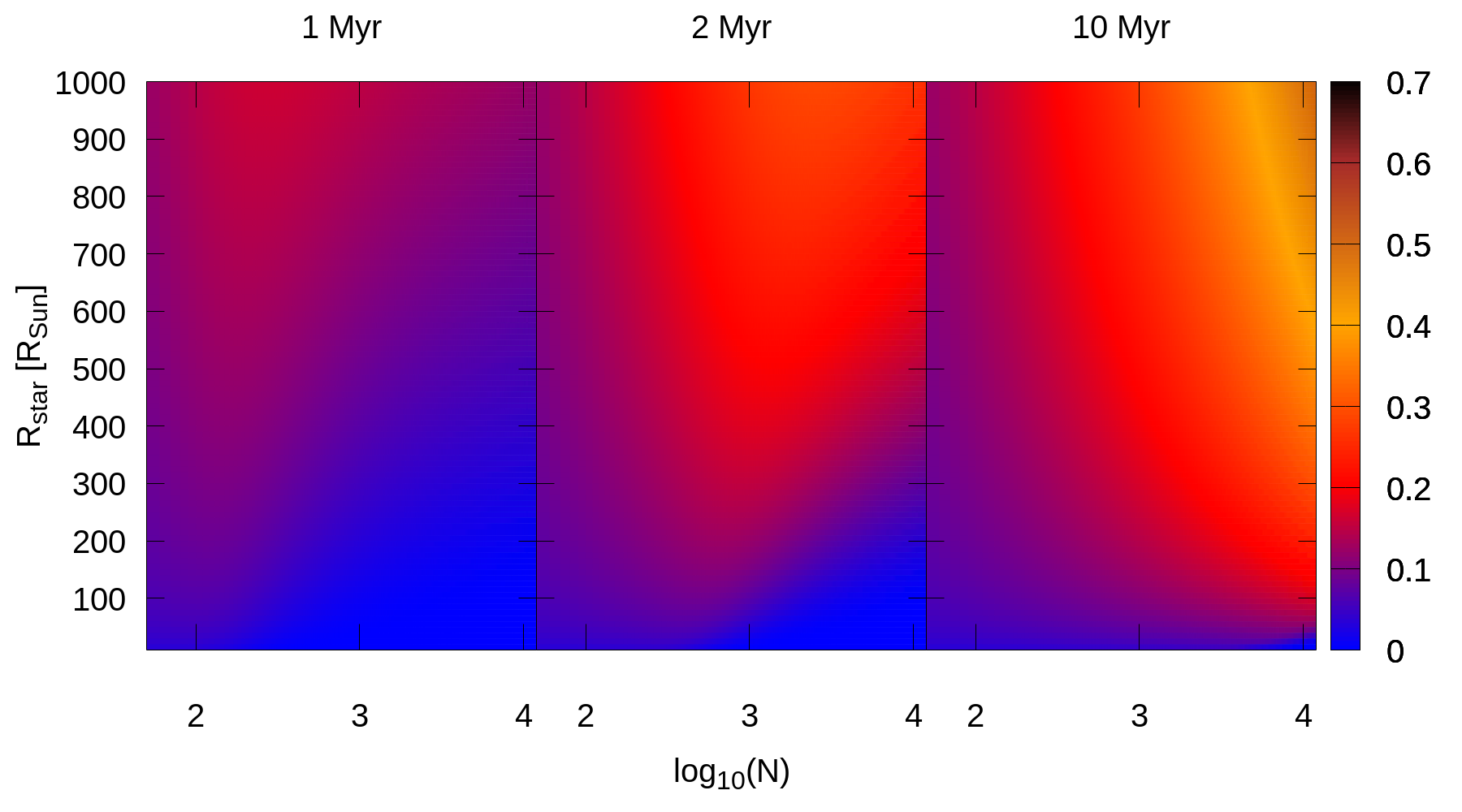}
      \caption{Fraction of stars $f_{\rm merger}$ that merge with the central star in a cluster with a total stellar mass of $M_{\rm stars}=10^4$~M$_{\odot}$ and a virial radius of $R_{\rm v}=0.14$~pc after 1~Myr (left panel), 2~Myr (middle panel), and 10~Myr (right panel) for different combinations of the number of stars $N$ and the stellar radius $R_{\rm star}$ calculated using our model presented in Eq.(\ref{eq:Nmerger_total}). We can see that when the collisions are maintained for short periods of time, that is, 1~Myr, a massive object with $\sim$ $f_{\rm merger}$~$M_{\rm stars}=0.15 \times 10^4$~M$_\odot$ = 1\,500~M$_\odot$ is formed in the cluster center only when $N$ is small (given that $t_{\rm delay}$ becomes larger than the integration time for large $N$) or large $R_{\rm star}$. If the time limit is long, that is, 10~Myr, most of the clusters will form stars with $\sim$ 0.3 $M_{\rm stars}=3\,000$~M$_\odot$.}
         \label{fig:fmerger_no_gas}
   \end{figure*}

\begin{figure*}
   \centering
   \includegraphics[width=\hsize]{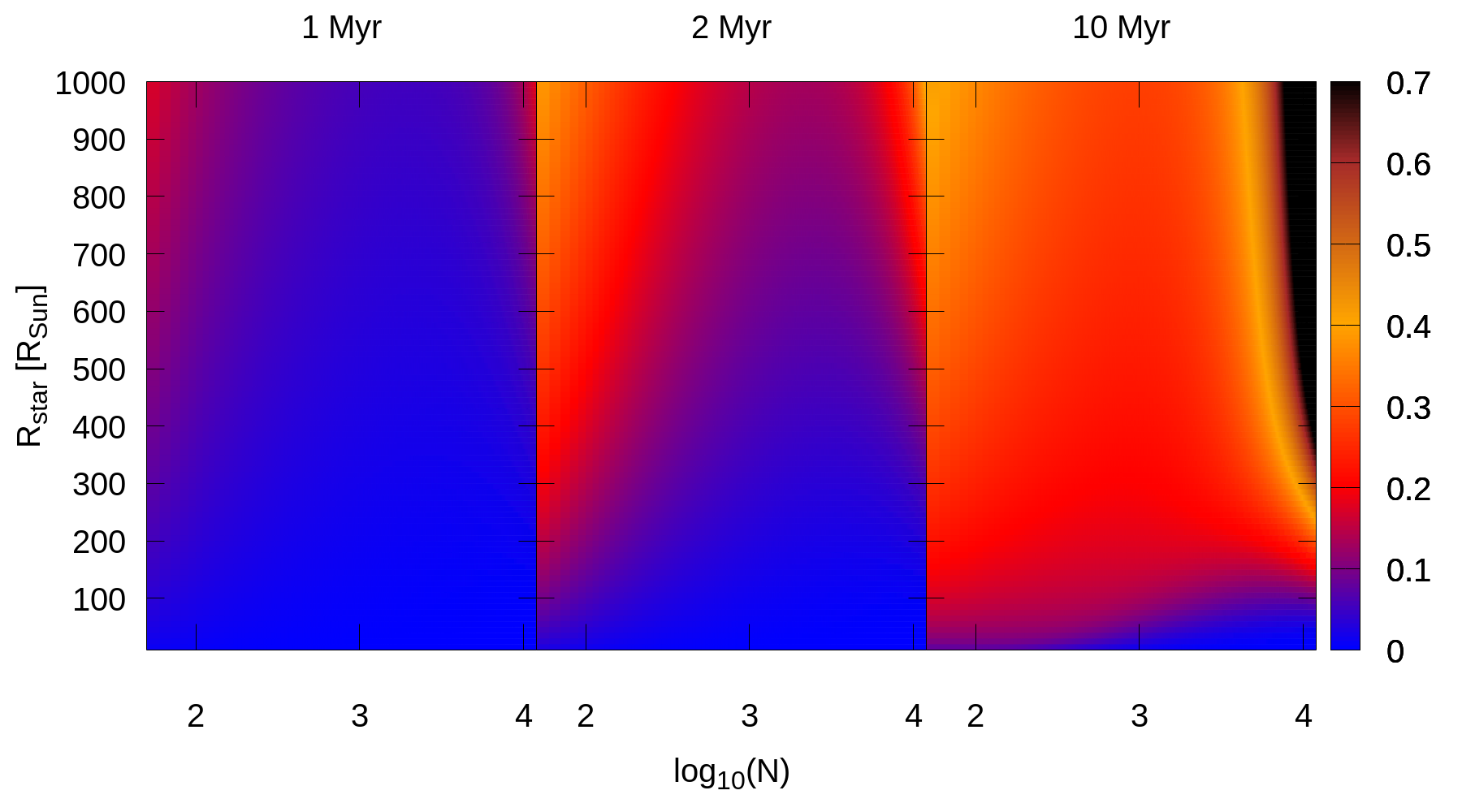}
      \caption{Fraction of stars $f_{\rm merger}$ that merge with the central star in a cluster with a total stellar mass of $M_{\rm stars}=10^4$~M$_{\odot}$, a virial radius of $R_{\rm v}=0.14$~pc, and an external potential with a mass of $M_{\rm ext}=M_{\rm stars}$ after 1~Myr (left panel), 2~Myr (middle panel), and 10~Myr (right panel) for different combinations of the number of stars $N$ and the stellar radius $R_{\rm star}$ calculated using our model presented in Eq.(\ref{eq:Nmerger_total}). We can see that when the collisions are maintained for short periods of time, that is, 1~Myr, an object that is not so massive with $\sim$ $f_{\rm merger}$~$M_{\rm stars}=0.1 \times 10^4$~M$_\odot$ = 1000~M$_\odot$ is formed in the cluster center only when $N$ is very small (given that $t_{\rm delay}$ becomes larger than the integration time for large $N$) or when $R_{\rm star}$ is large. If the time limit is 2~Myr (middle panel), we expect that clusters with a small $N$ experience core collapse and form a massive central star. However, if the number of stars is large and $R_{\rm star}$ is also large, given the higher velocity of the stars in those clusters compared to clusters without the external potential, there is a larger probability for hyperbolic mergers to occur before core collapse, which we see on the right-hand side of middle panel.
      If the time limit is long, that is, 10~Myr, most of the clusters form stars with $\sim$ 0.3-0.4~$M_{\rm stars}=3\,000-4\,000$~M$_\odot$ and even 0.7~$M_{\rm stars}$=7\,000~M$_\odot$ if $N$ is large enough.}
         \label{fig:fmerger_gas}
   \end{figure*}

Now we can estimate the number of mergers with the central star and thus the mass of this star up to a time $t$ for clusters with different initial conditions, including the presence of an external potential that is comparable to the total stellar mass in the cluster. This can be used as a very simplified model of a star cluster, which still contains gas during or shortly after the process of star formation or a nuclear star cluster. In Fig.~\ref{fig:fmerger_no_gas}, we show the expected fraction $f_{\rm merger}$ of the stars that merge with a single central object in a cluster without an external potential. In Fig.~\ref{fig:fmerger_gas}, this is shown for clusters in an external potential, for a broad combination of $N$ and $R_{\rm star}$ after 1, 2, and 10~Myr assuming a cluster with $M_{\rm stars}=10^4$~M$_\odot$, a virial radius of $R_{\rm v}=0.14$~pc, and a mass of the external potential $M_{\rm ext}=M_{\rm stars}$ for the second case. We used Eq.(\ref{eq:Nmerger_total}) for these calculations along with the parameters from Eqs.(\ref{eq:A_no_gas}), (\ref{eq:tdelay_no_gas}), (\ref{eq:t_duration_no_gas}), and (\ref{eq:D_no_gas}) to obtain the results presented in Fig.~\ref{fig:fmerger_no_gas}, and Eqs.(\ref{eq:A_gas}), (\ref{eq:tdelay_gas}), (\ref{eq:t_duration_gas}), and (\ref{eq:D_gas}) to obtain the results presented in Fig.~\ref{fig:fmerger_gas}. Our model gives the total number of stars $N_{\rm merger,total}$ that merge with the central object in a given time interval, and thus assuming that all stars are equal during the time at which most of the collisions occur, we can estimate the mass of the central object as $M_{\rm max}=N_{\rm merger,total} \overline{m}=N_{\rm merger,total} M_{\rm stars}/N = f_{\rm merger} M_{\rm stars}$.
Our results indicate that when collisions are maintained for short periods of time only, that is, 1~Myr, then the clusters without an external potential form more massive central stars than clusters with an external potential (see left panels of Figs.~\ref{fig:fmerger_no_gas} and \ref{fig:fmerger_gas}). This is due to the fact that the external potential dramatically increases the relaxation time of the cluster, as shown in Eq.(\ref{eq:trelax_q}). In fact the relaxation time depends on the ratio of external potential mass to stellar mass $q=M_{\rm ext}/M_{\rm stars}$ as $t_{\rm relax} \propto q^4$.

As the time limit increases, we see that in both clusters with and without an external potential a more massive star emerges in the center. However, even more massive stars form in the cluster without an external potential for most values of $N$. However, there is an important difference in clusters that include a background potential, which is that the potential keeps the cluster compact as previously found by \cite{Leigh2014}, and thus most of the cluster mass is able to eventually sink to the center at later times. This effect is visible in the left part of the middle panel in Figs.~\ref{fig:fmerger_no_gas} and \ref{fig:fmerger_gas}. We also note that for large $N$ and $R_{\rm star}$ for clusters in a background potential and before core collapse, we may also expect mergers with the central star,not binary mergers but hyperbolic, given that in those cases the cross section for collisions is very large and the probability of hyperbolic collisions increases with stellar velocity, that is, if there is an external potential. Finally when the time limit is very long, that is, 10~Myr, we may expect final masses for the central object in the order of 0.3-0.4~$M_{\rm stars}$ for most clusters without an external potential and 0.4-0.7 for clusters in a background potential because the compactness of the cluster is maintained and collisions still occur at an approximately constant rate after the stage of runaway growth of the central star.

\section{Implications for primordial clusters}
\label{sec:implications_for_primordial_clusters}
In the following, we explore the implications of our results with respect to primordial star clusters, including both embedded and gas free. 
Here we particularly distinguish the case of standard Pop.~III clusters as expected in a typical minihalo with about $10^6$~M$_{\odot}$.
\subsection{Standard Pop.~III clusters (minihalos)}
For a typical Pop.~III star cluster, we assume a mass of 1\,000~M$_{\odot}$, which is consistent with a baryon fraction of about 10\% and a star formation efficiency on the order of 1\% in a 10$^6$~M$_\odot$
 minihalo. We adopt a virial radius for the cluster on the order of 0.1~pc, which is consistent with the results from simulations and semi-analytic models \citep{Clark2011b,Clark2011a,Greif2011,Greif2012,Latif2013a,Latif2015}. We use a stellar radius of 100~R$_\odot$, which is characteristic of primordial protostars with accretion rates on the order of $10^{-3}$~M$_\odot$~yr$^{-1}$ \citep{Hosokawa2012}. The crossing time of the cluster then is 0.015~Myr. The number of stars that can be expected in such a cluster is uncertain, but here we adopt an estimate of about $N=100$. Thus, a stellar mass of $m_{\rm star} = 10$~M$_\odot$.
 Using the relations that we found, we expect two collisions to occur within 1~Myr which correspond to a final mass of 30~M$_{\odot}$. If we assume now a lifetime of 10~Myr, we can expect a total of six collisions with the central star which correspond to a final mass of 70~M$_\odot$.
 
 We consider now an embedded and accreting Pop.~III star cluster with a stellar mass of 1\,000~M$_\odot$, a gas mass of 1\,000~M$_\odot$, a virial radius of 0.1~pc, and a mean stellar radius of 100~R$_\odot$. The crossing time of the cluster, according to Eq.(\ref{eq:tcross}), corresponds to 0.007~Myr. We also adopt a number of stars $N=100$. Then our model predicts that no collisions occur within 1~Myr, but for a lifetime of 10~Myr we expect a total of 15 collisions and hence a central star with 160~M$_\odot$.
 We expect the lifetime of a massive primordial star to be in between this range, depending on the precise mass, the amount of rotation, and the effects that the collisions may have on the stellar evolution \citep{Maeder2012}. We therefore find that a moderate enhancement can be achieved within a normal cluster.
 We note that the values given here are the expected mean number of mergers. Individual clusters can deviate from these, both towards lower and higher fractions of mergers, potentially including clusters with zero mergers. This is especially the case for typical Pop.~III clusters where the number of collisions can be expected to be comparable to the mean value because it is a chaotic collisional dynamical process.
 
 \subsection{Massive primordial clusters (atomic cooling halos)}
 As a next step, we address now the potential impact of collisions in a more massive atomic cooling halo with a total mass of 10$^8$~M$_\odot$. Under the right conditions and in particular if the cooling on larger scales is regulated by atomic hydrogen \citep{Latif2014}, a rather massive cluster of 10$^4$~M$_\odot$ can form, which is then exposed to larger accretion rates on the order of 10$^{-1}$~M$_\odot$~yr$^{-1}$. We assume that the cluster consists of an initial number $N=1\,000$ of stars, the virial radius is $R_{\rm v}=0.14$~pc, and that the stellar radii are somewhat enhanced compared to the standard Pop.~III cluster due to the higher accretion rates, with a typical radius of about 300~R$_\odot$. The crossing time of the cluster according to Eq.(\ref{eq:tcross}) with $M_{\rm tot}=10^4$~M$_{\odot}$ is then 0.0078~Myr.
 Using our derived model we expect about 51 collisions in 1~Myr, and about 168 within 10~Myr with a single central star. We again expect the realistic lifetime of the resulting massive star to be in between these extreme cases. In the case of an atomic cooling halo, we thus conclude that a considerable enhancement is possible as a result of stellar mergers. If we take the mean stellar mass, then the expected masses are 520~-~1\,690~M$_\odot$ after 1 and 10~Myr, respectively.
 
 If we now consider the same cluster configuration but during the embedded phase, assuming a total mass in gas of 10$^4$~M$_\odot$, the crossing time of the cluster according to Eq.(\ref{eq:tcross}) with $q=M_{\rm ext}/M_{\rm stars}=2$ is then 0.0039~Myr, and we expect a total of 16 collisions to occur within 1~Myr and a total of 199 mergers within 10~Myr. Again, by taking the mean stellar, mass this corresponds to a mass of 170~-~2\,000~M$_\odot$ within 1 and 10~Myr, respectively.
 Also the reported values here correspond to a mean, and there can be deviations to lower and higher merger fractions. Regardless, we expect the number of collisions and hence the final mass to remain within the same order of magnitude.

\section{Discussion}
\label{sec:discussion}
We modeled the runaway growth of stars through mergers in the center of dense star clusters by including an analytic external potential in our $N$-body simulations. Our model relies on computing four parameters that depend on the number of stars $N$ and their stellar radii $R_{\rm star}$. These parameters are the delay time $t_{\rm delay}$, the duration time for the collision process $t_{\rm duration}$, and two normalization factors $A$ and $D$. We then used these four parameters to integrate a Gaussian function and estimate the mass of the central object up to a time $t$. Although a Gaussian fit is a good tool to estimate $t_{\rm delay}$, $t_{\rm duration}$, and $A$ (see Fig.~\ref{fig:Gausian_example}), there is a deviation from this function when we include an external potential. In fact, when we compare the mass growth of the central object in clusters with and without an external potential (see, e.g. Figs.~\ref{fig:N1k_Mg0_R100_merger} and \ref{fig:N1k_Mg1_R100_merger}), we see a delay for clusters in an external potential and an initially less dramatic growth until the onset of mergers of stars in binary systems. This growth continues after the runaway growth and is seen as tails of a Gaussian in a plot of $N_{\rm mergers}$ versus time (see top panel of Figs.~\ref{fig:Mmax_no_gas} and \ref{fig:Mmax_gas}). This later growth is associated to hyperbolic mergers, that is, mergers of stars that are not bound by gravity, and this type of merger becomes important when including an external potential because the cluster remains compact for longer periods of times \citep{Leigh2014}. The onset of collisions is delayed in simulations with an external potential relative to simulations without it.

We also have found a formula for the delay time when we include an external potential $t_{\rm delay,ext}=t_{\rm delay}+t_{{\rm delay},M_{\rm ext}=M_{\rm stars}}$, which is basically the same expression for clusters without the external potential plus an additional term $t_{{\rm delay},M_{\rm ext}=M_{\rm stars}}$  presented in Eq.(\ref{eq:tdelay_gas}). We present this parameter in this way so that we can investigate in the future if this can be used in a more general way as a function of the mass of the external field $M_{\rm ext}$ and if it is related to a longer relaxation time by means of its dependence on $M_{\rm ext}$ as suggested by Eq.(\ref{eq:trelax_q}).

The model derived here can be used as a first approximation to obtain the number of stellar mergers in embedded star clusters or to understand the effects of an external potential on the formation of massive merger products, which might also be important for the modelling of nuclear star clusters. Future research in this field will employ hydrodynamic modeling of the gas, which may prevent a large delay in the onset of runaway growth for the central star due to gas accretion  and dissipative effects via star-gas or binary-gas interactions. These simulations also need to include realistic mass-radius relations for accreting protostars when aiming for a realistic modeling of the first star clusters in the Universe.
By performing these suites of N-body simulations we have covered a large parameter space, and this allows us to provide hints as to which part of this parameter space the future, more sophisticated simulations should be focused on and the subsequent physics that becomes relevant and often even dominant in these regimes.

Our simulations including mergers adopt a virial radius of $R_{\rm v} =0.14$~pc and a total stellar mass of $M_{\rm stars}=10^4$~M$_\odot$, but our results can be rescaled for different sizes and masses by means of the crossing time. In this work, we only consider for the initial conditions equal mass and equal radii stars but allow for the mass and radius to vary due to the mergers according to Eqs.~(\ref{eq:merger_new_mass}) and (\ref{eq:merger_new_radius}), and in this sense our simulations represent an effective model where an average stellar radius is adopted over the period of time considered. This effective stellar radius should correspond to the typical radius when the majority of mergers are expected to occur.

Our model still lacks a more realistic stellar population, namely an initial mass function (IMF), which would naturally lead to mass segregation and subsequent mergers of the most massive stars in the cluster center. Moreover, we have not included the dissipative effects expected for star-gas interactions in our simulations that include an external potential, which could lead to the formation of more binary systems and hence more mergers \citep{Leigh2014}. In this sense, we do not expect a significant variation in the final mass of the most massive stars if the total mass in the most massive stars is comparable to the final mass obtained in our models, around 30~-~2\,000~M$_\odot$. Additionally, including an IMF, three-body encounters and two-body relaxation would cause the ejection of the smallest stars in the cluster, which could be an important source of stellar relics from the first star clusters in the early Universe.
Another important way to proceed is a better modeling of the stellar mergers, in particular the mass loss during the stellar collisions which may be up to 25\% of the total mass \citep{Gaburov2010} and the collision product may even receive a kick velocity > 10~km~s$^{-1}$. More recent work on the effects of mass loss and its effect on runaway growth of a central star in a cluster shows that considering 5\% mass loss in every collision produces objects which are 20~-~40\% less massive at the end of the runaway growth compared to models without mass loss \citep{Glebbeek2009,Alister2019}.
Thus regarding this, the results presented here should be taken as an upper limit on the mass of the merger product.

\section{Conclusions}
\label{sec:conclusions}
In this study, we have performed a set of 374 $N$-body simulations of dense, virialized star clusters and clusters in the center of an external potential including stellar mergers in order to understand the effects of the background force on the formation of massive stars and derive a model to estimate the mass enhancement of these objects in embedded star clusters.
We find that the presence of an external potential delays the overall evolution of the star cluster and the formation of a massive central star through runaway collisions due to the increased kinetic energy of the stars, which in turn increases the relaxation time. However, the merger products become more massive (if the collision process is maintained for a long time) given that these clusters expand less than clusters without an external potential and so more stars are able to merge with the central star even after the process of runaway growth. 

We also find that the increased velocity dispersion for star clusters in an external potential boosts the number of hyperbolic mergers , that is, mergers of stars that are not part of a binary or triple system, with up to 50-60\% of the total number of mergers being hyperbolic mergers. Whereas in clusters without an external potential, this percentage is around 30\%. This is due to the fact that in presence of a background potential, the clusters remain more compact \citep{Leigh2013,Leigh2014} and hyperbolic mergers still may occur. Stellar ejections are highly suppressed in clusters in the center of an external potential.

We find a set of equations that can be used to estimate the mass of the merger product at different times for both clusters with and without an external potential, and we present an example of such calculations in Figs.~\ref{fig:fmerger_no_gas}, \ref{fig:fmerger_gas}, and in Sec.\ref{sec:implications_for_primordial_clusters} proving that if the process is not interrupted for long periods of time ($\geq 10$~Myr), the external potential enhances the mass of the central object by a factor of $\sim$2. However, if the process is interrupted at early times ($\sim$1~Myr), the clusters in a background potential produce objects only half as massive compared to objects formed in clusters without an external potential. When applied to Pop.~III star clusters we find, for standard clusters formed in a minihalo, a moderate enhancement for the mass of the most massive star in the range of 10~-~160~M$_\odot$ within 1 and 10~Myr in embedded clusters. Whereas for a massive Pop.~III star cluster that formed in an atomic cooling halo, we find that the mass of the most massive star lies in the range of 170~-~2\,000~M$_\odot$ in embedded clusters within 1 and 10~Myr.

\begin{acknowledgements}
We thank the anonymous referee for the useful comments. BR thanks Sverre Aarseth for his help with the code NBODY6. BR also thanks Conicyt for funding through (CONICYT-PFCHA/Mag\'isterNacional/2017-22171385), (CONICYT-PFCHA/Doctorado acuerdo bilateral DAAD/62180013), financial support from DAAD (funding program number 57451854). 
DRGS acknowledges support of Fondecyt Regular 1161247 and through Conicyt PIA ACT172033. BR and DRGS acknowledge financial support through Anillo (project number ACT172033).
MF acknowledges funding through Fondecyt Regular No.1180291.
NWCL gratefully acknowledges the support of a Fondecyt Iniciacion \#11180005.
\end{acknowledgements}

\bibliography{gaspot}
\bibliographystyle{bibtex/aa}

\begin{appendix} 
\section{Modification of the half-mass relaxation time-scale}
\label{sec:appendix1}
We have seen that when we include an external potential in a star cluster, the overall evolution seems to be delayed, in particular the core collapse times-scale rapidly increases with the mass of the external potential. Until now, we have used the usual half-mass relaxation time-scale defined in Eq.(\ref{eq:tcc_spitzer}) \citep{Spitzer1987}. However, this is probably not adequate when we include an external potential because in that case, considering a virialized cluster, the root mean square (rms) velocity of the stars is higher due to the extra force exerted by the potential and this certainly modifies the time-scale on which stellar encounters are going to modify the higher velocity of the stars.
In order to account for this effect, here, we derive a modified relaxation time-scale with the aid of a new parameter $q=M_{\rm ext}/M_{\rm stars}$.
We begin with the usual derivation of the relaxation time assuming virial equilibrium, a similar derivation can be found in \cite{BinneyTremaine1987}:
\begin{equation}
\label{eq:virial_equi}
    2T+U=0.
\end{equation}
We note that $T=\frac{1}{2}M_{\rm stars}<v^2>$ is the kinetic energy and $|U|=\frac{GM}{2R_{\rm v}}$ is the potential energy. In this case, $M_{\rm stars}$ is the total mass in stars, $<v^2>$ is the mean velocity squared of the stars, $M=M_{\rm stars}+M_{\rm ext}$ is the total mass, that is, the mass in stars plus the mass of the external potential $M_{\rm ext}$, and $R_{\rm v}$ is the virial radius of the cluster, which in this case is the same virial radius for the external potential. Now we write the total mass as a function of $q$:
\begin{equation*}
    M=M_{\rm stars} + M_{\rm ext},
\end{equation*}
\begin{equation*}
    M=M_{\rm stars} \left( 1+\frac{M_{\rm ext}}{M_{\rm stars}} \right),
\end{equation*}
\begin{equation}
\label{eq:total_massq}
    M=M_{\rm stars}(1+q).
\end{equation}

We begin now with a derivation for the change of the velocity of a star due to an encounter with another star. The force that one star feels due to the other star is (see Fig.\ref{fig:derivation_trelax_1}):

   \begin{figure}[h]
   \centering
   \includegraphics[width=\hsize]{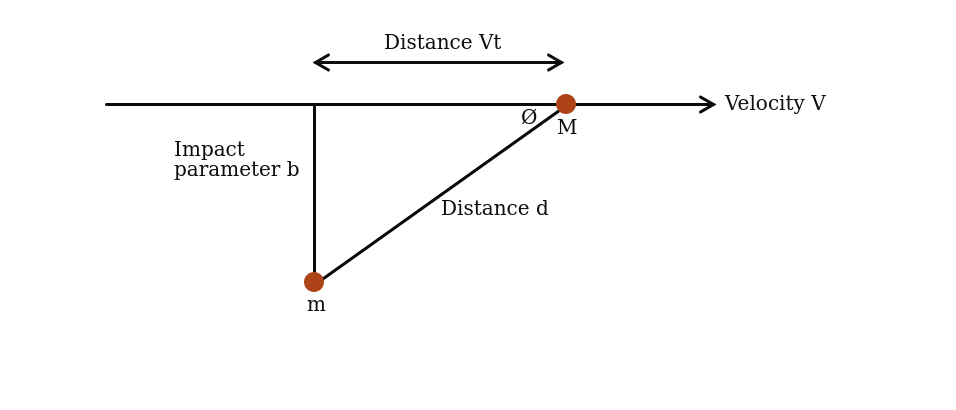}
      \caption{Trajectory of a star with mass M that passes close to another star of mass m, which causes a small deflection on the velocity vector V.}
         \label{fig:derivation_trelax_1}
   \end{figure}

\begin{equation}
    F=\frac{GMm}{d^2}=\frac{GMm}{b^2+V^2t^2},
\end{equation}
where $M$ is the mass of one star and $m$ is the mass of the other star, and $d$ is the impact parameter or distance of closest approach. Now we assume that the velocity change $\Delta V$ of the star with mass $M$ in the direction parallel to $V$ is small, and only the perpendicular component  of the velocity is changed due to the perpendicular force:
\begin{equation}
    F_{\perp} = F \sin{\phi}=F\times \frac{b}{d}=\frac{GMmb}{(b^2+V^2t^2)^{3/2}},
\end{equation}
\begin{equation}
\label{eq:f_perp1}
    F_{\perp} = M\frac{dV_{\perp}}{dt}.
\end{equation}
Next we want to know the time at which the perpendicular velocity changes by an amount $\Delta V \sim V$, which is the relaxation time-scale. Therefore, we integrate Eq.(\ref{eq:f_perp1}) so the final perpendicular velocity is:
\begin{eqnarray*}
    \Delta V&=&\displaystyle \int_{-\infty}^{\infty} \frac{dV_{\perp}}{dt} dt,\\
    &=&\displaystyle \int_{-\infty}^{\infty} \frac{F_{\perp}}{M} dt,\\
    &=&\displaystyle \int_{-\infty}^{\infty} \frac{Gmb}{(b^2+V^2t^2)^{3/2}}dt,\\
    &=& \frac{2Gm} {bV}.
\end{eqnarray*}
As the star receives many deflections in different directions, we are interested in the mean value of the squares of the velocity kicks $\langle \Delta V_\perp^{2}\rangle$ that can be found integrating all the small deflections as:
\begin{equation*}
    \langle \Delta V_\perp^2\rangle = \displaystyle \int_{b_{\rm min}}^{b_{\rm max}} \left( \frac{2Gm}{bV}\right)^2 dN,
\end{equation*}
where $dN$ is the expected number of encounters that occurs in a time $t$ between impact parameters $b$ and $b+db$ for a star with typical velocity $V$, this is:
\begin{equation*}
    dN=n\times Vt \times 2\pi b db,
\end{equation*}
where $n$ is the number density of stars, then,

\begin{eqnarray}
\langle \Delta V^2 \rangle&=&\displaystyle \int_{b_{\rm min}}^{b_{\rm max}} nVt \left( \frac{2Gm}{bV}\right)^2 2\pi b db, \nonumber  \\ 
&=&\frac{8\pi G^2 m^2 nt}{V}\displaystyle \int_{b_{\rm min}}^{b_{\rm max}} \frac{db}{b}, \nonumber \\ 
\label{eq:delta_v_squared}
&=&\frac{8\pi G^2 m^2 nt}{V}\ln{ \left( \frac{b_{\rm max}}{b_{\rm min}} \right)}.
\end{eqnarray}
After enough time, the perpendicular velocity of one star grows to its original speed, this time is the relaxation time-scale that we can derive using Eq.(\ref{eq:delta_v_squared}):
\begin{equation}
\label{eq:trelax_usual}
   t_{\rm relax}= \frac{V^3}{8\pi G^2 m^2 n \ln{ \left( \frac{b_{\rm max}}{b_{\rm min}} \right)} }.
\end{equation}

The term $\frac{b_{\rm max}}{b_{\rm min}}$ is often written as $\Lambda$ and is the ratio of the size of the system to the "strong encounter distance" $b_{\rm min}=2Gm/V^2$, which is the distance at which an encounter with another star would result in a 90 degree deflection. This ratio $\frac{b_{\rm max}}{b_{\rm min}}$, for a system of identical $N$ stars is found to be $\frac{b_{\rm max}}{b_{\rm min}}=\gamma N$  \citep{Spitzer1987}, with $\gamma=0.4$. This definition of the relaxation time-scale is useful to estimate the core collapse time, which typically occurs between 15-20 half-mass relaxation times. The problem comes when we include an analytic external potential and evolve the cluster at the center of this potential. In this case, we have an extra force acting on the stars, and therefore a virialized cluster under the influence of this external potential has a larger potential energy and also a larger kinetic energy compared to a cluster in which there is not an external potential. This should increase the relaxation time-sale as the velocity of the stars increases and the number and masses of stars is kept constant. We now derive the relaxation time for a cluster under the influence of an external potential, assuming that this potential follows the same mass distribution of the stars and with the same virial radius. In Recalling the condition for virial equilibrium from Eq.(\ref{eq:virial_equi}) and the total mass for the cluster from Eq.(\ref{eq:total_massq}) we find that:
\begin{eqnarray}
2T+U&=&0, \nonumber\\
2\left(\frac{1}{2} NmV^2\right)-\frac{GM_{\rm tot}^2}{R_{\rm v}}&=&0, \nonumber\\
M_{\rm stars}V^2&=&\frac{GM^{2}_{\rm stars}(1+q)^2}{R_{\rm v}}, \nonumber\\
V&=&\sqrt{\frac{GM_{\rm stars}}{R_{\rm v}}}(1+q), \nonumber 
\end{eqnarray}
where $V$ is the rms velocity of the stars, $N$ is the number of
stars, m is the mass of a single star, $M=M_{\rm stars}+M_{\rm ext}$ is the total mass, that is, the mass of the stars plus the mass of the external potential, and $q=\left(1+\frac{M_{\rm ext}}{M_{\rm stars}} \right)$ and $R_{\rm v}$ is the virial radius of the system.
When $q=0$, we get the typical velocity for stars in a cluster which is in virial equilibrium:
\begin{equation}
    V=\sqrt{\frac{GM_{\rm stars}}{R_{\rm v}}},
\end{equation}
but for a cluster in virial equilibrium and with a background potential, the typical velocity is modified as:
\begin{eqnarray}
    V_{\rm ext}&=&\sqrt{\frac{GM_{\rm stars}}{R_{\rm v}}}(1+q), \nonumber \\
\label{eq:velocity_on_q}
    V_{\rm ext}&=&V(1+q),
\end{eqnarray}
where $V_{\rm ext}$ is the rms velocity of the stars in the presence of an external potential and $V$ is the rms velocity of stars without a background potential.

The relaxation time is often expressed as a function of the crossing time $t_{\rm cross}$ of the cluster, which is simply the time it takes for a star with the typical velocity $V$ to cross the system:
\begin{equation}
\label{eq:tcross_usual}
    t_{\rm cross}=\frac{R_{\rm v}}{V}, 
\end{equation}
then, combining Eqs.(\ref{eq:trelax_usual}) and (\ref{eq:tcross_usual}) the relaxation time for clusters without an external potential is:
\begin{equation}
    t_{\rm relax}=\frac{V^4}{8\pi G^2 m^2 n R_{\rm v} \ln{ \left( \frac{b_{\rm max}}{b_{\rm min}} \right)} } t_{\rm cross}. 
\end{equation}

Now, if we include an external potential, $q\neq 0$, and in substituting Eqs.(\ref{eq:velocity_on_q}) for (\ref{eq:tcross_usual}), the crossing time for a cluster in an external potential is:
\begin{eqnarray}
t_{\rm cross,ext}&=&\frac{R_{\rm v}}{V(1+q)}, \nonumber \\
\label{eq:trelax_on_q_tcros}
t_{\rm cross,ext}&=&\frac{t_{\rm cross}}{(1+q)}. 
\end{eqnarray}

Subsequently the relaxation time for clusters in an external potential must be:
\begin{eqnarray}
    \label{eq:trelax,ext_aux}
    t_{\rm relax,ext}&=&\frac{V_{\rm ext}^4}{8\pi G^2 m^2 n R_{\rm v} \ln{ \left( \frac{b_{\rm max}}{b_{\rm min}} \right)} } t_{\rm cross,ext}. \nonumber \\
\end{eqnarray}

Now we replace Eq.(\ref{eq:velocity_on_q}) for Eq.(\ref{eq:trelax,ext_aux}) and we find:

\begin{eqnarray}
    \label{eq:trelax,ext}
    t_{\rm relax,ext}&=&\frac{V^4 (1+q)^4}{8\pi G^2 m^2 n R_{\rm v} \ln{ \left( \frac{b_{\rm max}}{b_{\rm min}} \right)} } t_{\rm cross,ext},
\end{eqnarray}


Now in recalling that from the usual definition of the relaxation time we can define the half-mass relaxation time for a cluster of $N$ equal mass stars as \citep{Spitzer1987}:
\begin{equation*}
    t_{\rm rh}=0.138\frac{N}{\ln{(\gamma N)}}t_{\rm cross}
\end{equation*}
then, for comparison purposes, we define the half-mass relaxation time-scale for clusters in an external potential as:
\begin{equation}
\label{eq:trelax_for_ext_pot}
    t_{\rm rh,ext}=0.138\frac{N(1+q)^4}{\ln{(\gamma N)}} t_{\rm cross,ext},
\end{equation}
with $q=\frac{M_{\rm ext}}{M_{\rm stars}}$ and $\gamma=0.4$ for equal mass stars.

\section{Evolution of clusters in a background potential}
\label{sec:appendix2}
In this section, we describe in more detail the results of our simulations that do not include stellar collisions and which are listed in Table~\ref{tab:sim_no_mergers}. We show the evolution of the clusters with and without an external potential, and we find a delay in the overall evolution when increasing the mass of the external potential.

\begin{figure}
   \centering
   \includegraphics[width=\hsize]{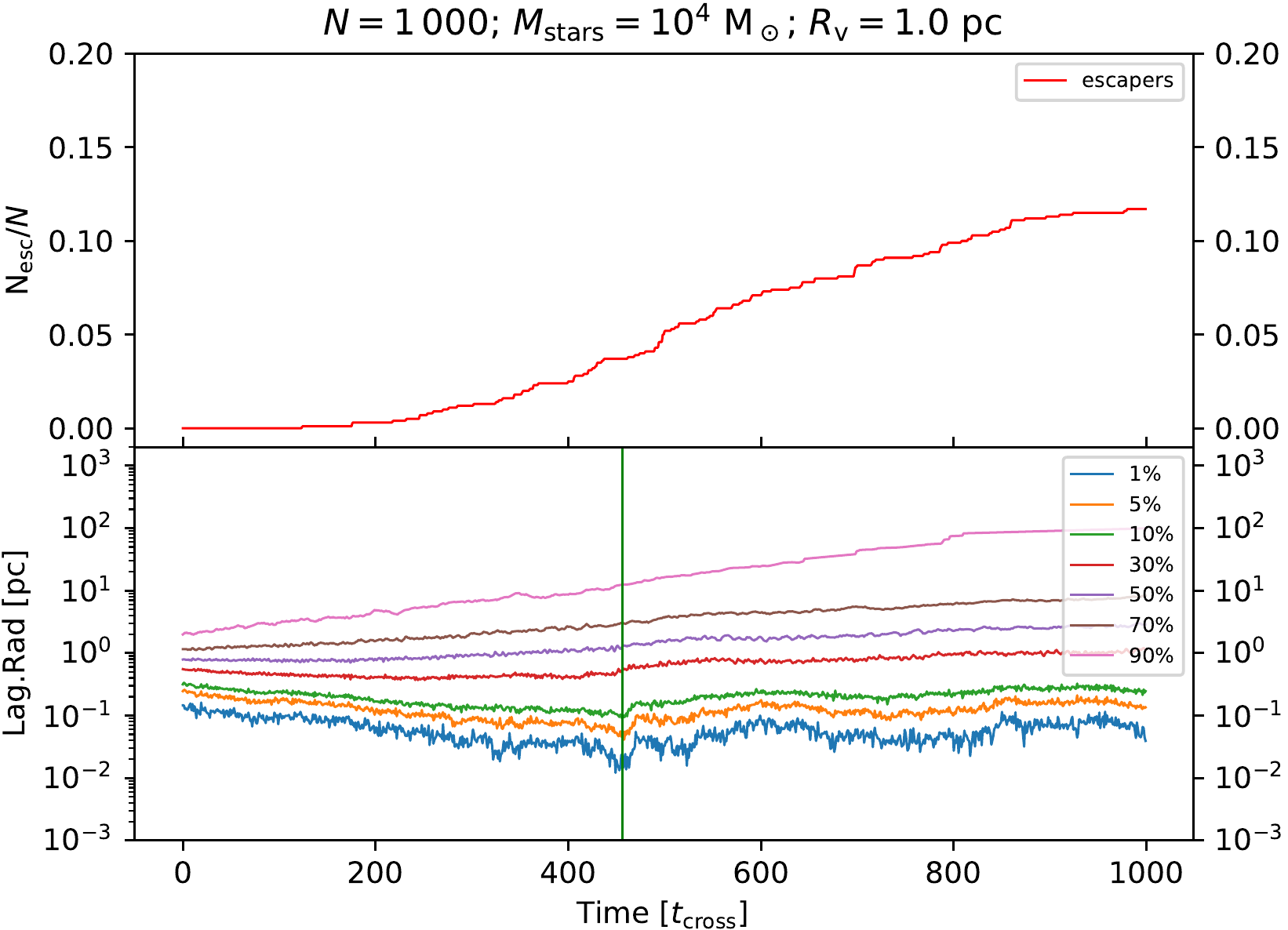}
      \caption{Evolution of a cluster with $N=1\,000$ stars, total mass $M_{\rm stars}=10^4$~M$_\odot$ and $R_{\rm v}=1.0$ pc. The top panel shows the fraction of stars that were ejected from the cluster. The bottom panel shows the Lagrangian radius and the vertical green line marks the moment of core collapse. The time is presented in units of the crossing time of the cluster.}
         \label{fig:N1k_no_merger_Mg0}
   \end{figure}

All of the clusters evolve toward core collapse, which we found upon visual inspection by looking for the first drop and subsequent rise in the 10\% Lagrangian radius. The core collapse occurs for the clusters without a background potential at 456 $t_{\rm cross}=19.78$~$t_{\rm rh}$ for the cluster with $N=1\,000$ (see Fig.~\ref{fig:N1k_no_merger_Mg0}), and for the cluster with $N=10\,000$ at 2\,306~$t_{\rm cross}= 13.85$~$t_{\rm rh}$ (see Fig.~{\ref{fig:N1k_no_merger_Mg0}}).

   \begin{figure}
   \centering
   \includegraphics[width=\hsize]{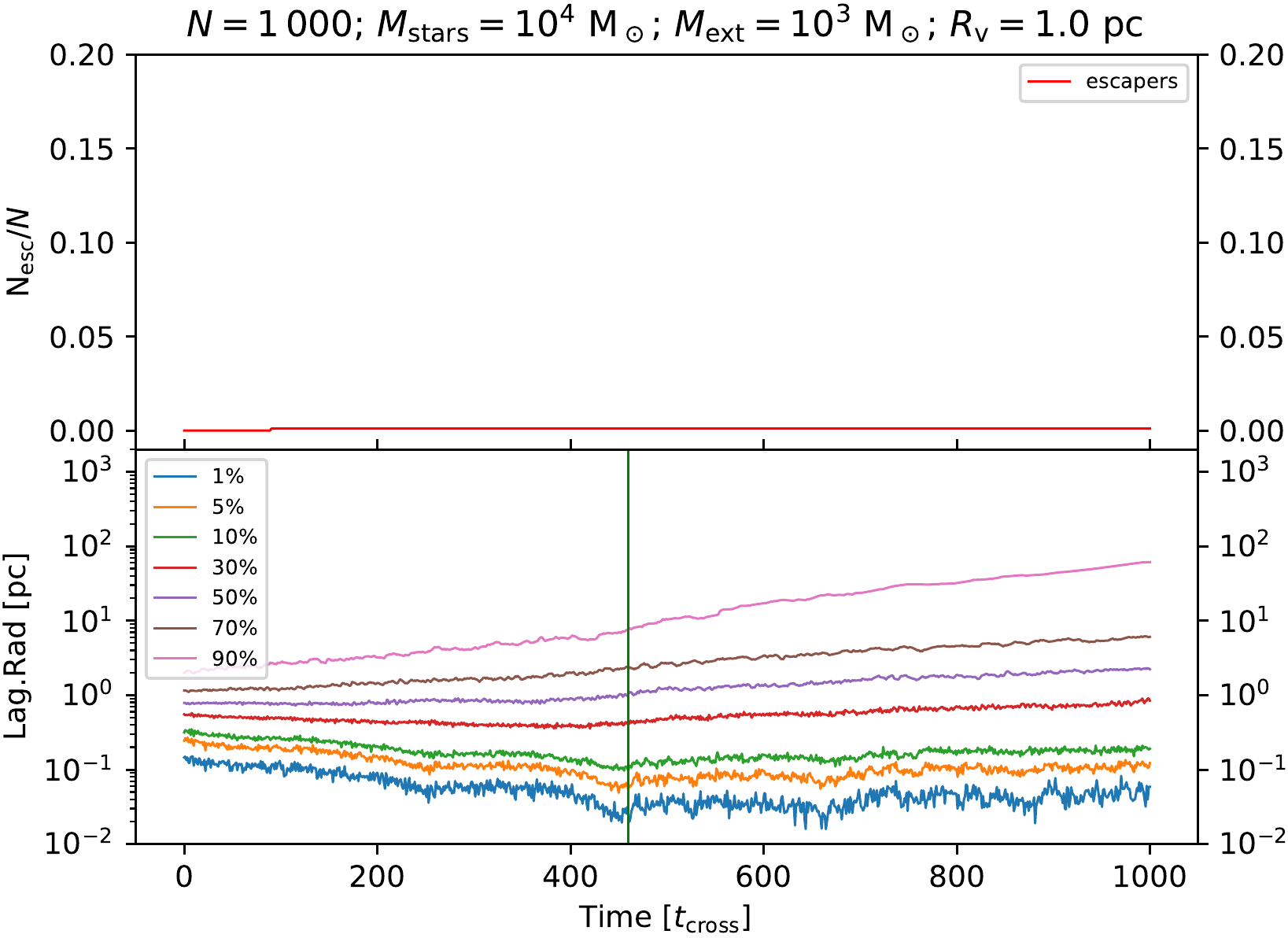}
      \caption{Evolution of a cluster with $N=1\,000$ stars, total mass $M_{\rm stars}=10^4$~M$_\odot$ and $R_{\rm v}=1.0$ pc in a background potential with mass $M_{\rm ext}=10^3$~M$_\odot$. The top panel shows the fraction of stars that were ejected from the cluster. The bottom panel shows the Lagrangian radius and the vertical green line marks the moment of core collapse. The time is presented in units of the crossing time of the cluster.}
         \label{fig:N1k_no_merger_Mg01}
   \end{figure}

      \begin{figure}
   \centering
   \includegraphics[width=\hsize]{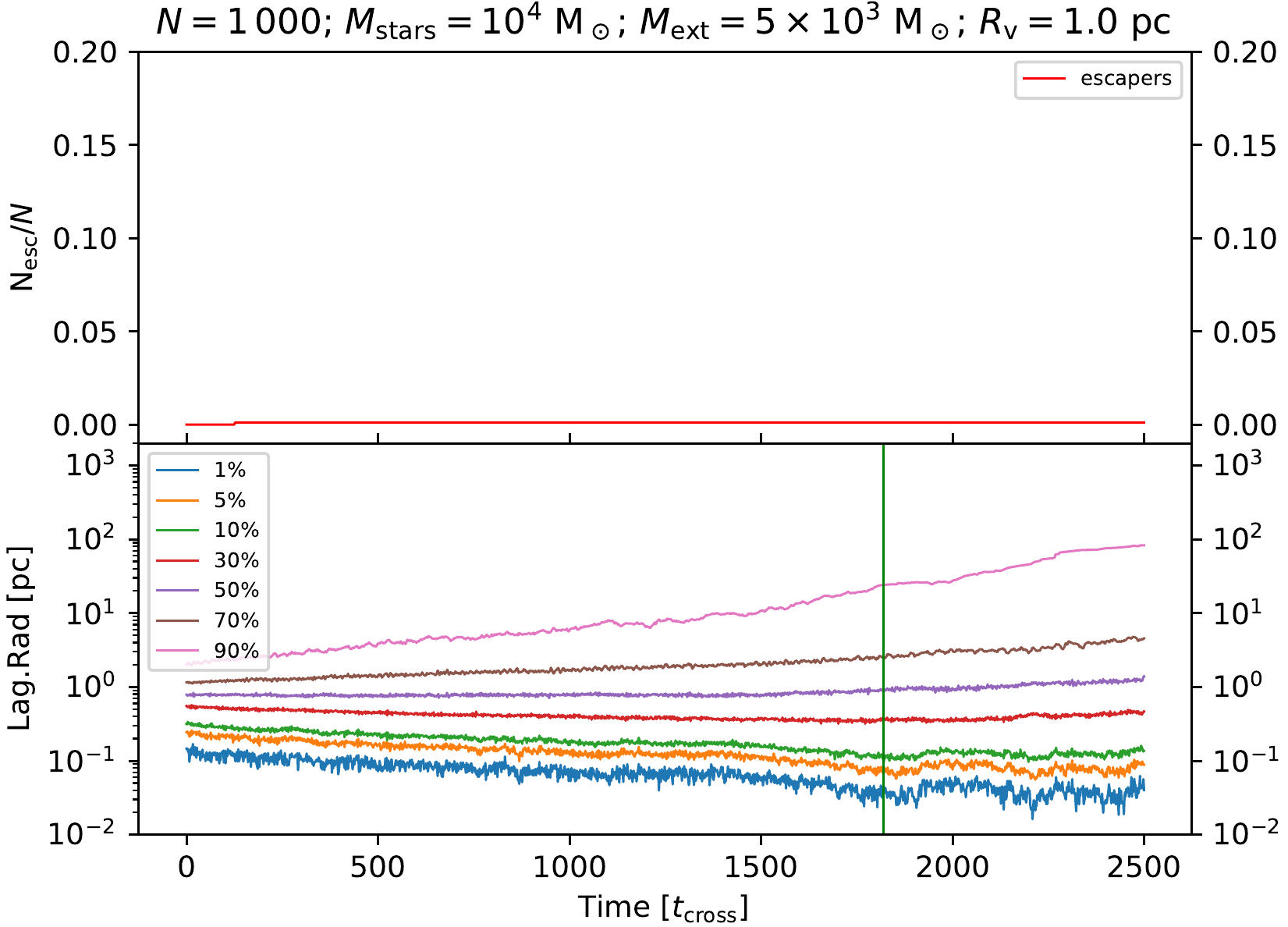}
      \caption{Evolution of a cluster with $N=1\,000$ stars, total mass $M_{\rm stars}=10^4$~M$_\odot$, $R_{\rm v}=1.0$ pc in a background potential with mass $M_{\rm ext}=5\times 10^3$~M$_\odot$. The top panel shows the fraction of stars that were ejected from the cluster. The bottom panel shows the Lagrangian radius and the vertical green line marks the moment of core collapse. The time is presented in units of the crossing time of the cluster.}
         \label{fig:N1k_no_merger_Mg05}
   \end{figure}

         \begin{figure}
   \centering
   \includegraphics[width=\hsize]{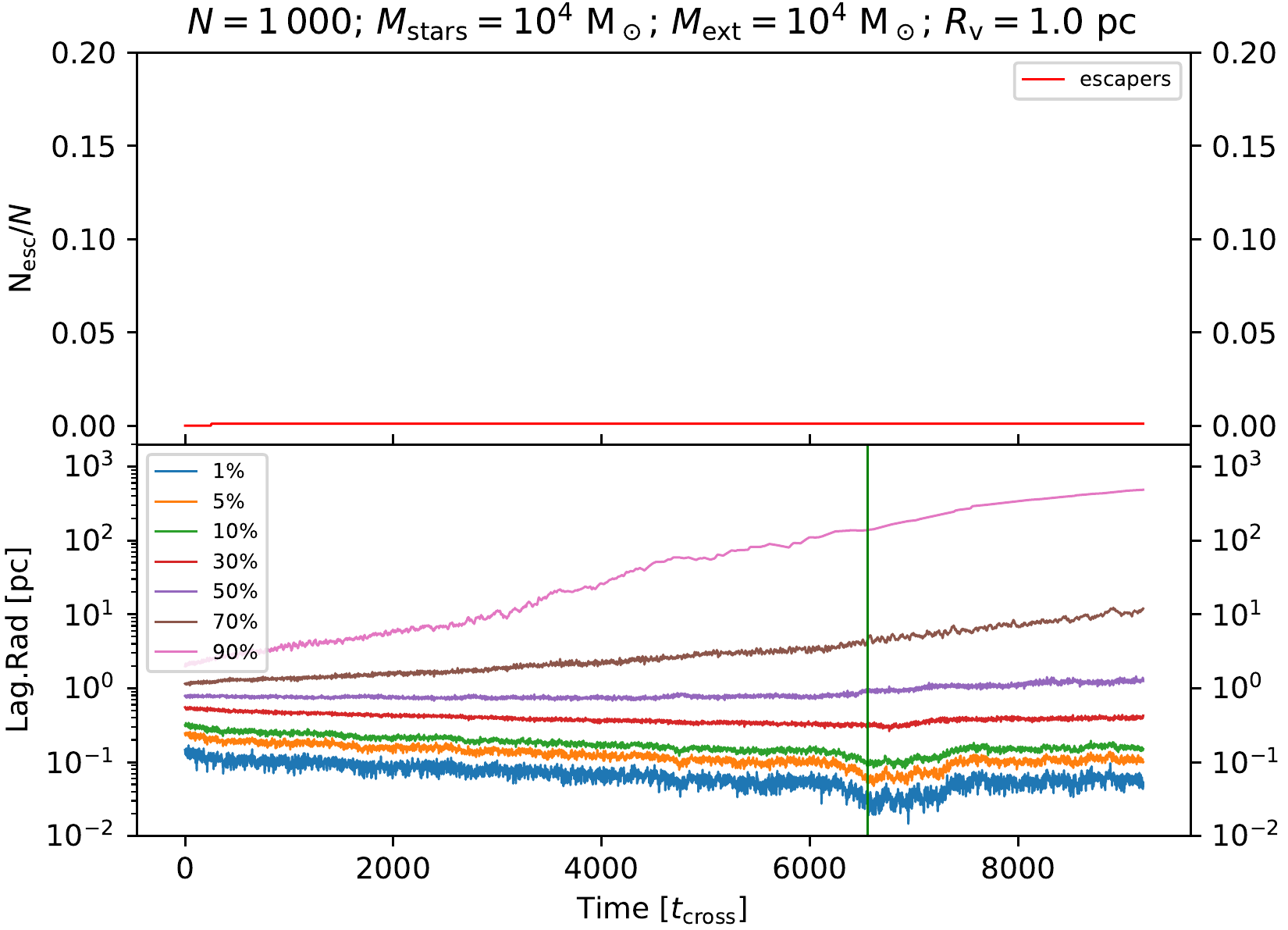}
      \caption{Evolution of a cluster with $N=1\,000$ stars, total mass $M_{\rm stars}=10^4$~M$_\odot$, $R_{\rm v}=1.0$ pc in a background potential with mass $M_{\rm ext}=10^4$~M$_\odot$. The top panel shows the fraction of stars that were ejected from the cluster. The bottom panel shows the Lagrangian radius and the vertical green line marks the moment of core collapse. The time is presented in units of the crossing time of the cluster.}
         \label{fig:N1k_no_merger_Mg1}
   \end{figure}

As the clusters evolve toward core collapse, the outer parts continually expand which leads, along with the onset of stellar ejections, to the evaporation of the clusters. We also note that the onset of ejections occur just after core collapse when binary systems form in the core and three body interactions lead to the ejection of stars. An ejection occurs if the distance from the center of mass of the cluster to a star is larger than 20 times the virial radius $R_{\rm v}$ and if the total energy of the star is greater than zero, that is, E$_{\rm star} = \rm E_{\rm kin} + |E_{\rm pot}| > 0$ with E$_{\rm kin}$ and E$_{\rm pot}$ being the kinetic and potential energy, respectively. These results are shown in Figs.~\ref{fig:N10k_no_gas} and \ref{fig:N1k_no_merger_Mg0}.

When we include a background potential with a low mass compared to the total mass in stars, $M_{\rm ext}= 0.1 \times M_{\rm stars}$ in our simulations with $N=1\,000$, the core collapse time is 460~$t_{\rm cross}$, which is similar to the core collapse time for the cluster without the external potential in terms of the crossing time of the cluster. However, this cluster show less expansion and only one star is ejected (see Fig.~\ref{fig:N1k_no_merger_Mg01}).

For the case when the mass of the background potential is half the mass in stars, $M_{\rm ext} = 0.5 \times$~$M_{\rm stars}$ the cluster evolves toward core collapse, which now occurs at 1\,820~$t_{\rm cross}=16$~$t_{\rm rh}$ (calculated with Eq.(\ref{eq:trelax_for_ext_pot})). The cluster again shows less expansion than the cluster without the external potential (see Fig.\ref{fig:N1k_no_merger_Mg0}) and only one single star is ejected.

For the case when the mass of the external potential is the same as the total mass in stars $M_{\rm ext}= M_{\rm stars}$, the cluster also evolves toward core collapse. However, this is now delayed now until 6\,553~$t_{\rm cross}$ as shown by the green vertical line in Fig.\ref{fig:N1k_no_merger_Mg1}.

\begin{figure}
   \centering
   \includegraphics[width=\hsize]{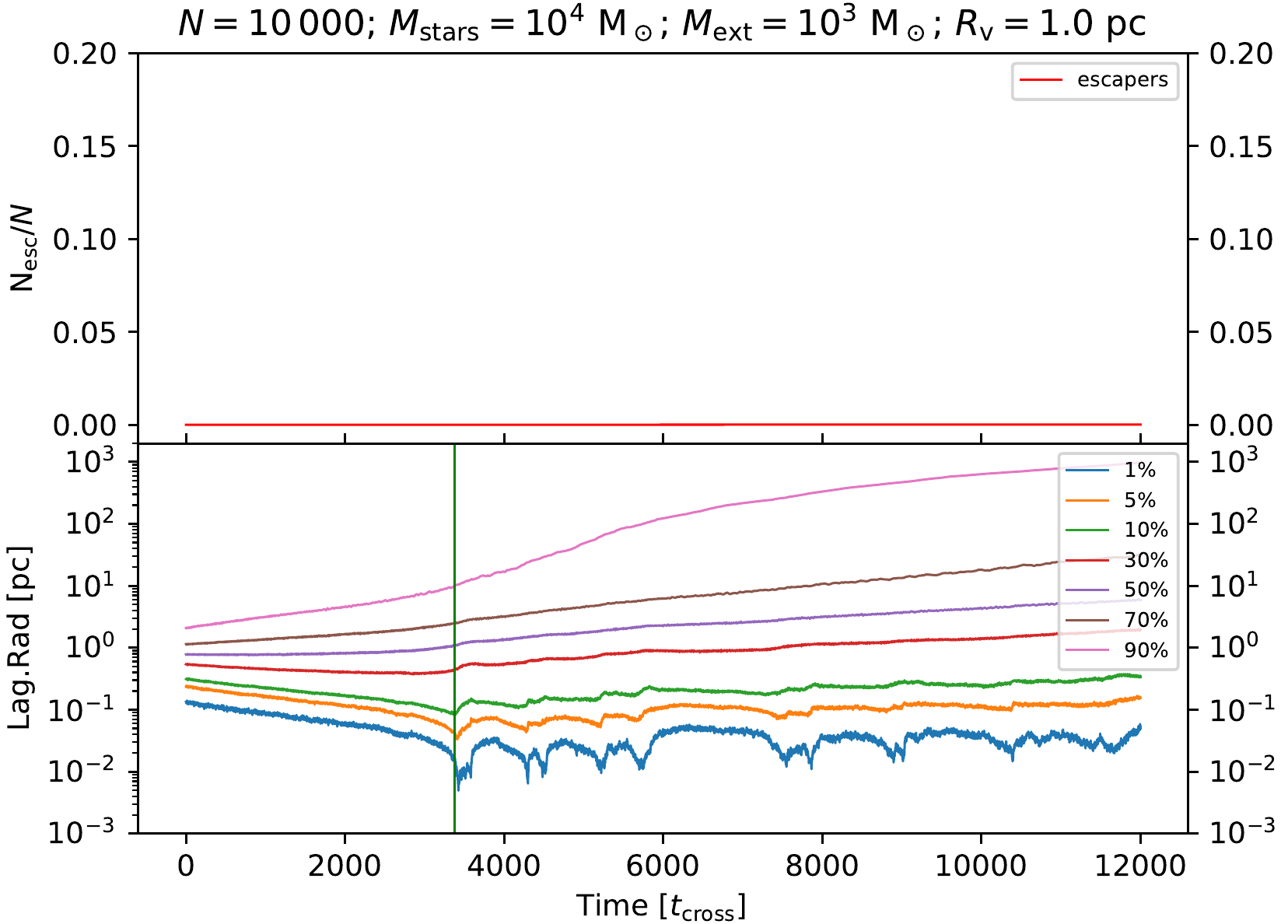}
      \caption{Evolution of a cluster with $N=10\,000$ stars, total mass $M_{\rm stars}=10^4$~M$_\odot$ and $R_{\rm v}=1.0$ pc under the influence of a background potential with $M_{\rm ext}=10^3$~$M_{\rm ext}$. The top panel shows the fraction of stars that were ejected from the cluster, which is negligible since only one star escaped the cluster. The bottom panel shows the Lagrangian radius and the vertical green line marks the moment of core collapse. The time is presented in units of the crossing time of the cluster.}
         \label{fig:N10k_no_merger_Mg01}
   \end{figure}

When the number of stars is $N=10\,000$, the behavior is essentially the same when we include a background potential; the global evolution is delayed due to the increased velocity of the stars. First, we present the case when the mass of the external potential is low $M_{\rm ext}= 0.1\times M_{\rm stars}$ compared to the total mass in stars. The cluster also evolves toward core collapse, which now occurs at 3\,377~$t_{\rm cross}$ as indicated with a green vertical line in the bottom panel of Fig.~\ref{fig:N10k_no_merger_Mg01}, which is $\sim$14~$t_{\rm rh}$ (we used Eq.(\ref{eq:trelax_for_ext_pot}) to calculate the half mass relaxation time). The mean density inside the 10\% Lagrangian radius at this moment is of 4.4$\times 10^5$~M$_\odot$~pc$^{-3}$, which is higher than the mean density at core collapse for the cluster without a background potential. Moreover the cluster in general shows less expansion when including the external potential.

\begin{figure}
   \centering
   \includegraphics[width=\hsize]{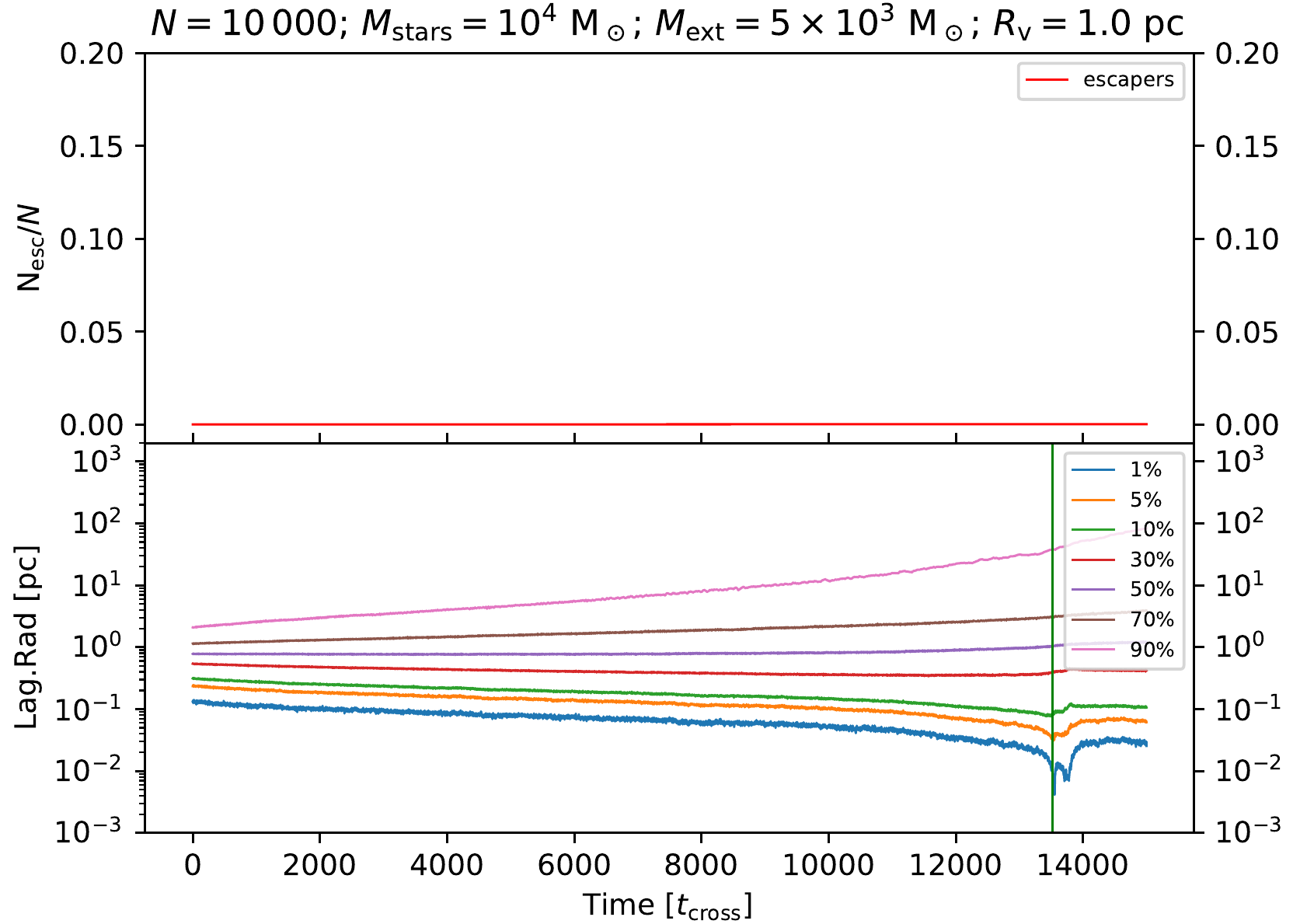}
      \caption{Evolution of a cluster with $N=10\,000$ stars, total mass $M_{\rm stars}=10^4$~M$_\odot$ and $R_{\rm v}=1.0$ pc under the influence of a background potential with $M_{\rm ext}=5 \times 10^3$~$M_{\rm ext}$. The top panel shows the fraction of stars that were ejected from the cluster, which is negligible since only one star escaped the cluster. The bottom panel shows the Lagrangian radius and the vertical green line marks the moment of core collapse. The time is presented in units of the crossing time of the cluster.}
         \label{fig:N10k_no_merger_Mg05}
   \end{figure}

When the mass of the external potential is half the total mass in stars, that is, $M_{\rm ext}=0.5 \times M_{\rm stars}$, the core collapse is delayed until 13\,519~$t_{\rm cross}$ as marked by the green vertical line in Fig.~\ref{fig:N10k_no_merger_Mg05} which is 16~$t_{\rm rh}$ (we used Eq.(\ref{eq:trelax_for_ext_pot}) again to calculate the half mass relaxation time) and during this moment the mean density inside the 10\% Lagrangian radius is of 5.5$\times 10^5$~M$_\odot$~pc$^{-3}$. Our simulation is not long enough to see the expansion of the outer parts. however we do expect even less expansion than for the cluster with $M_{\rm ext}=0.1\times M_{\rm stars}$ and again until this time only 1 star has been ejected from the cluster, this was also found in \cite{Leigh2013} and \cite{Leigh2014}.

\end{appendix}

\end{document}